\documentclass[11pt,draftcls,onecolumn]{IEEEtran}

\usepackage{amsbsy,epsfig,epsf,url,citesort}

\usepackage{graphicx}
\usepackage{amssymb,amsmath,dsfont,amsfonts}
\usepackage{epstopdf}
\usepackage{cite}
\usepackage{color}
\usepackage{boxedminipage}
\usepackage{listings}
\usepackage{subfigure}
\usepackage{caption2}
\usepackage{float}
\usepackage{cases}

\title{Technical Report: Observability \\with Random Observations}

\author{Borhan M. Sanandaj$\text{i,}^{*}$ Michael B. Waki$\text{n,}^{\diamond}$ and Tyrone L. Vincen$\text{t}^{\diamond}$%
\thanks{\textsuperscript{$\star$}B. M. Sanandaji is with the Department of Electrical Engineering and Computer Sciences at the University of California, Berkeley, CA 94720, USA. Email: sanandaji@eecs.berkeley.edu.~\textsuperscript{$\diamond$}M. B. Wakin and T. L. Vincent are with the Department of Electrical Engineering and Computer Science at the Colorado School of Mines, Golden, CO 80401, USA. Email: \{mwakin, tvincent\}@mines.edu. Preliminary versions of portions of this work appeared in~\cite{wakin2010observability}. This work was partially supported by AFOSR Grant FA9550-09-1-0465, NSF Grant CCF-0830320, DARPA Grant HR0011-08-1-0078, and NSF Grant CNS-0931748.}
}

\def\real    { \mathbb{R} }

\newtheorem{theorem}{Theorem}
\newtheorem{cor}{Corollary}
\newtheorem{lemma}{Lemma}
\newtheorem{definition}{Definition}

\newcommand{\eps}{\epsilon}
\newcommand{\bitem}{\begin{itemize}}
\newcommand{\eitem}{\end{itemize}}

\newcommand{\beqn}{\begin{equation}}
\newcommand{\eeqn}{\end{equation}}
\newcommand{\balign}{\begin{align}}
\newcommand{\ealign}{\end{align}}

\def\ss{\vspace*{-3mm}}

\newcommand{\norm}[1]{\| #1 \|}

\newcommand{\Prob}[1]{{\bf P}\left\{#1\right\}}

\def \ok {{\mathcal{O}_{\Omega}}}
\def \ck {{\mathcal{C}_{\Omega}}}
\def \ak {{\mathcal{A}_{\Omega}}}
\def \km {{\widetilde{M}}}
\def \kn {{\widetilde{N}}}

\newcommand{\cut}[1]{}

\newcommand{\vc}[1]{\boldsymbol{#1}}

\def\real    { \mathbb{R} }

\usepackage{acronym}
\acrodef{i.i.d.}{independent and identically distributed}
\acrodef{LTI}{Linear Time-Invariant}
\acrodef{RIP}{Restricted Isometry Property}
\acrodef{SVD}{Singular Value Decomposition}
\acrodef{CS}{Compressive Sensing}
\acrodef{DSP}{Digital Signal Processing}

\acrodef{FIR}{Finite Impulse Response}
\acrodef{LTI}{linear time-invariant}
\acrodef{DFT}{Discrete Fourier Transform}
\acrodef{JL}{Johnson-Lindenstrauss}
\acrodef{ROC}{Receiver Operating Curve}
\acrodef{NP}{Neyman-Pearson}
\acrodef{CoM}{Concentration of Measure}
\acrodef{CSI}{Compressive System Identification}
\acrodef{CBD}{Compressive Binary Detection}

\begin{document}

\maketitle

\begin{abstract}

Recovery of the initial state of a high-dimensional system can require a large number of measurements. In this paper, we explain how this burden can be significantly reduced when randomized measurement operators are employed. Our work builds upon recent results from Compressive Sensing (CS). In particular, we make the connection to CS analysis for random
block diagonal matrices.
By deriving Concentration of Measure (CoM) inequalities, we show that the observability matrix satisfies the Restricted Isometry Property (RIP) (a sufficient condition for stable recovery of sparse vectors) under certain conditions on the state transition matrix. For example, we show that if the state transition matrix is unitary, and if independent, randomly-populated measurement matrices are employed, then it is possible to uniquely recover a sparse high-dimensional initial state when the total number of measurements scales \emph{linearly} in the sparsity level (the number of non-zero entries) of the initial state and logarithmically in the state dimension. \cut{This is in fact a significant reduction in the sufficient total number of measurement for correct initial state recovery as compared to traditional observability theory.}We further extend our RIP analysis for scaled unitary and symmetric state transition matrices. We support our analysis with a case study of a two-dimensional diffusion process.
\end{abstract}




\section{Introduction}
\label{sec:intro}

In this paper, we consider the problem of recovering the initial state of a high-dimensional system from
compressive measurements~(i.e., we take fewer measurements than the system dimension).


\subsection{Measurement Burdens in Observability Theory}

Consider an $N$-dimensional discrete-time linear dynamical system described by the state equation\footnote{The results of this paper also apply directly to systems described by a state equation of the form
\begin{equation*}
\begin{array}{rcl}
\vc{x}_k &=& A \vc{x}_{k-1} + B\vc{u}_k\\[1mm]
\vc{y}_k &=& C_k \vc{x}_k+ D\vc{u}_k,
\end{array}
\end{equation*}
where $\vc{u}_k \in \real^P$ is the input vector at sample time $k$ and $B \in \real^{N \times P}$ and $D \in \real^{M \times P}$ are constant matrices. 
Indeed, initial state recovery is independent of $B$ and $D$ when it is assumed that the input vector $\vc{u}_k$ is known for all sample times $k$.
}
\begin{equation}
\begin{array}{rcl}
\vc{x}_k &=& A \vc{x}_{k-1} \\[1mm]
\vc{y}_k &=& C_k \vc{x}_k,
\end{array}
\label{eq:sys}
\end{equation}
where $\vc{x}_k \in \real^N$ represents the state vector at time $k \in \{0,1,2,\dots\}$, $A \in \real^{N \times N}$ represents the state transition matrix, $\vc{y}_k \in \real^M$ represents a set of measurements (or ``observations'') of the state at time $k$, and $C_k \in \real^{M \times N}$ represents the measurement matrix at time $k$. (Observe that the number of measurements at each sample time is $M$.)
For any finite set $\Omega \subset \left\{0, 1, 2, 3, \dots \right\}$, define the {\em generalized observability matrix} as
\begin{equation}
\mathcal{O}_{\Omega} := \left[ \begin{array}{c} C_{k_0}A^{k_0} \\ C_{k_1} A^{k_1} \\ \vdots \\ C_{k_{K-1}} A^{k_{K-1}} \end{array} \right] \in \real^{MK \times N},
\label{eq:ok}
\end{equation}
where $\Omega = \left\{k_0, k_1, \dots, k_{K-1}\right\}$ contains $K$ observation times. Note that this definition extends the traditional definition of the observability matrix by allowing arbitrary time samples in (\ref{eq:ok}) and matches the traditional definition when $\Omega = \left\{0, 1, \dots, K-1\right\}$. The primary use of observability theory is in ensuring that a state (say, an initial state $\vc{x}_0$) can be recovered from a collection of measurements $\left\{\vc{y}_{k_0}, \vc{y}_{k_1}, \dots, \vc{y}_{k_{K-1}}\right\}$. In particular, defining $\vc{y}_{\Omega} := \left[\vc{y}^T_{k_0}~~\vc{y}^T_{k_1}~~\cdots~~\vc{y}^T_{k_{K-1}} \right]^T \in \real^{MK}$,
we have
\begin{equation}
\vc{y}_{\Omega} = \mathcal{O}_{\Omega}\vc{x}_0.
\label{eq:matvec1}
\end{equation}

Although we will consider situations where $C_k$ changes with each $k$, we first discuss the classical case where $C_k = C$ ($C$ is assumed to have full row rank) for all $k$ and $\Omega = \left\{0,1,\dots,K-1\right\}$ (the observation times are consecutive). In this setting, an important and classical result~\cite{Chen99} states that a system described by the state equation (\ref{eq:sys}) is observable if and only if $\mathcal{O}_{\Omega}$ has rank $N$ (full column rank) where $\Omega = \left\{0,1,\dots,N-1\right\}$. One challenge in exploiting this fact is that for some systems, $N$ can be quite large. For example, distributed systems evolving on a spatial domain can have a large state space even after taking a spatially-discretized approximation.
In such settings, we might therefore require a very large total number of measurements ($MK$ with $K=N$) to identify an initial state, and moreover, inverting the matrix $\mathcal{O}_{\Omega}$ could be very computationally demanding.

This raises an interesting question: under what circumstances might we be able to infer the initial state of a system when $MK < N$?
We might imagine, for example, that the measurement burden could be alleviated in cases
when there is a model for the state $\vc{x}_0$ that we wish to recover. Alternatively, we may have cases where, rather than needing to recover $\vc{x}_0$ from $\vc{y}_{\Omega}$, we desire only to solve a much simpler inference problem such as a binary detection or a classification problem. In this paper, inspired by the emerging theory of Compressive Sensing (CS)~\cite{CompSenDon,CompSampCand,candes2008people}, we explain how such assumptions can indeed reduce the measurement burden and, in some cases, even allow recovery of the initial state when $MK < N$ and the system of equations (\ref{eq:matvec1}) is guaranteed to be underdetermined. More broadly, exploiting CS concepts in the analysis of sparse dynamical systems from limited information has gained much attention over the last few years in applications such as system identification~\cite{ohlsson2010segmentation,toth2011csi,sanandaji2012thesis,shah2012linear}, control feedback design~\cite{zhao2012stability}, and interconnected networks~\cite{sanandaji2011cti,pan2012reconstruction}.

\subsection{Compressive Sensing and Randomized Measurements}

The CS theory states that it is possible to solve certain rank-deficient sets of linear equations by imposing a model assumption on the signal to be recovered. In particular, suppose $\vc{y} = \Phi \vc{x}$ where $\Phi$ is an $\km \times N$ matrix with $\km < N$. Suppose also that $\vc{x} \in \real^N$ is $S$-sparse, meaning that only $S$ out of its $N$ entries are non-zero.\footnote{This is easily extended to the case where $\vc{x}$ is sparse in some transform domain.}
If $\Phi$ satisfies a condition called the~\ac{RIP} of order $2S$ for a sufficiently small isometry constant $\delta_{2S}$, then it is possible to {\emph{uniquely}} recover any $S$-sparse signal $\vc{x}$ from the measurements $\vc{y} = \Phi \vc{x}$ using a tractable convex optimization program known as $\ell_1$-minimization~\cite{CompSenDon,CompSampCand}. The RIP also ensures that the recovery process is robust to noise and stable in cases where $\vc{x}$ is not precisely sparse~\cite{CandesRIP}.\cut{Similar statements can be made for recovery using various iterative greedy algorithms~\cite{tropp2007signal,needell2009uniform,needell2009cosamp,dai2009subspace}.
}
In the following, we provide the definition of the \ac{RIP}.
\begin{definition}
A matrix $\Phi \in \real^{\km \times N}$ ($\km < N$) is said to satisfy the \ac{RIP} of order $S$ with isometry constant $\delta_S \in \left(0, 1\right)$ if
\begin{equation}
(1-\delta_S) \|\vc{x}\|^2_2 \le \| \Phi \vc{x}\|^2_2 \le (1+\delta_S) \|\vc{x}\|^2_2
\label{eq:rip}
\end{equation}
holds for all $S$-sparse vectors $\vc{x} \in \real^N$.
\label{def:rip}
\end{definition}

Observe that the \ac{RIP} is a property of a matrix and has a deterministic definition. However, checking whether the \ac{RIP} holds for a given matrix $\Phi$ is computationally expensive and is almost impossible when $N$ is large.
A common way to establish the \ac{RIP} for $\Phi$ is to populate $\Phi$ with random entries.
If $\Phi$ is populated with \ac{i.i.d.} Gaussian random variables having zero mean and variance $\frac{1}{\km}$, for example, then $\Phi$ will satisfy the \ac{RIP} of order $S$ with isometry constant $\delta_S$ with very high probability when $\km$ is 
proportional to $\delta^{-2}_SS \log \frac{N}{S}$~\cite{candes2008people,baraniuk2008simple,davenport2010thesis}.
This result is significant because it indicates that the number of measurements sufficient for correct recovery scales \emph{linearly} in the sparsity level $S$ and only \emph{logarithmically} in the ambient dimension $N$. Other random distributions may also be considered, including matrices with uniform entries of random signs.
Consequently, a number of new sensing hardware architectures, from analog-to-digital converters to digital cameras, are being developed to take advantage of the benefits of random measurements~\cite{duarte2008single,healySPmag,wakin2012non,yoo2012compressed}.

A simple way~\cite{baraniuk2008simple,DeVoreL1IO} of proving the \ac{RIP} for a randomized construction of $\Phi$ involves first showing that the matrix satisfies a \ac{CoM} inequality akin to the following.
\begin{definition}
A random matrix (a matrix whose entries are drawn from a particular probability distribution) $\Phi \in \real^{\km \times N}$ is said to satisfy the
Concentration of Measure (CoM) inequality if for any fixed signal $\vc{x} \in \real^N$ (not necessarily sparse) and any $\epsilon \in (0,\overline{\epsilon})$,
\begin{equation}
\Prob{\bigg{|} \| \Phi \vc{x} \|_2^2 - \|\vc{x}\|_2^2 \bigg{|} > \epsilon \|\vc{x}\|_2^2 } \leq 2 \exp \left\{ -\km f(\epsilon)\right\},
\label{eq:CoM_def}
\end{equation}
where $f(\epsilon)$ is a positive constant that depends on the isometry constant $\epsilon$, and $\overline{\epsilon} \le 1$ is some maximum value of the isometry constant for which the CoM inequality holds.
\label{def:concgauss}
\end{definition}

Note that the failure probability in~(\ref{eq:CoM_def}) decays exponentially fast in the number of measurements $\km$ times some constant $f\left(\epsilon\right)$ that depends on the isometry constant $\epsilon$. For most interesting random matrices, including matrices populated with \ac{i.i.d.} Gaussian random variables, $f(\epsilon)$ is quadratic in $\epsilon$ as $\epsilon \to 0$.

Baraniuk et al.~\cite{baraniuk2008simple} and Mendelson et al.~\cite{mendelson2008uniform} showed that a \ac{CoM} inequality of the form~(\ref{eq:CoM_def}) can be used to prove the \ac{RIP} for random compressive matrices.
This result is rephrased by Davenport~\cite{davenport2010thesis}.
\begin{lemma} {\em \cite{davenport2010thesis}}
Let $\mathcal{X}$ denote an $S$-dimensional subspace in $\real^N$.
Let $\delta_S \in (0,1)$ denote a distortion factor and $\nu \in (0,1)$ denote a failure probability, and suppose $\Phi$ is an $\km \times N$ random matrix that satisfies the \ac{CoM} inequality~(\ref{eq:CoM_def}) with 
$\km \geq \frac{S\log(\frac{42}{\delta_S})+ \log(\frac{2}{\nu})}{f(\frac{\delta_S}{\sqrt{2}})}$.
Then with probability at least $1-\nu$, for all $\vc{x} \in \mathcal{X}$,
\begin{equation*}
(1-\delta_S) \|\vc{x}\|^2_2 \le \| \Phi \vc{x}\|^2_2 \le (1+\delta_S) \|\vc{x}\|^2_2.
\end{equation*}
\label{lem:RIP_based_CoM}
\end{lemma}

Through a union bound argument (see, for example, Theorem~5.2 
in~\cite{baraniuk2008simple}) and by applying Lemma~\ref{lem:RIP_based_CoM} for all $N \choose S$ $S$-dimensional subspaces that define the space of $S$-sparse signals in $\real^N$, one can show that $\Phi$ satisfies the \ac{RIP} (of order $S$ and with isometry constant $\delta_S$) with high probability when $\km$ scales \emph{linearly} in $S$ and \emph{logarithmically} in $N$.
\cut{
Aside from connections to the \ac{RIP}, concentration inequalities such as the above can also be useful when solving other types of inference problems from compressive measurements. For example, rather than recovering a signal $\vc{x}$, one may wish only to solve a binary detection problem and determine whether a set of measurements $\vc{y}$ correspond only to noise (the null hypothesis $\vc{y} = \Phi(\mathrm{noise})$) or to signal plus noise ($\vc{y} = \Phi(\vc{x} + \mathrm{noise})$). When $\Phi$ is random, the performance of a compressive detector (and of other multi-signal classifiers) can be studied using concentration inequalities~\cite{davenport2010signal}, and in these settings it is not necessary to assume that $\vc{x}$ is sparse.
}
\subsection{Observability from Random, Compressive Measurements}

In order to exploit CS concepts in observability analysis, we consider scenarios where the measurement matrices $C_k$ are populated with random entries. Physically, such randomized measurements may be taken using the types of CS protocols and hardware mentioned above. Our analysis is therefore appropriate in cases where one has some control over the sensing process.

As is apparent from (\ref{eq:ok}), even with randomness in matrices $C_k$, the observability matrix $\ok$ contains some structure and cannot be simply modeled as a matrix populated with \ac{i.i.d.} Gaussian random variables and thus, existing results cannot be directly applied.
Our work builds on a recent paper by Park et al.\ in which \ac{CoM} inequalities are derived for random block diagonal matrices~\cite{park2011block}.
Our concentration results cover a large class of systems (not necessarily unitary) and initial states (not necessarily sparse), and apart from guaranteeing recovery of sparse initial states, other inference problems concerning $\vc{x}_0$, such as detection or classification of more general initial states and systems, can also be solved using random, compressive measurements, and the performance of such techniques can be studied using the \ac{CoM} bounds that we provide~\cite{davenport2010signal}.

Our \ac{CoM} results have important implications for establishing the \ac{RIP}.~Such \ac{RIP} results provide a sufficient number of measurements for exact initial state recovery when the initial state is known to be sparse a priori. The results of this paper show that under certain conditions on $A$
(e.g., for unitary, scaled unitary, and certain symmetric matrices $A$), the observability matrix $\ok$ will satisfy the
\ac{RIP} with high probability when the total number of
measurements $MK$ scales linearly in $S$ and logarithmically in $N$.
Before going into the details of the derivations and proofs, we first state in Section~\ref{sec:rip} our main results on establishing the \ac{RIP} for the observability matrix.
We then present in Section~\ref{sec:com} the \ac{CoM} results upon which the conclusions for establishing the RIP are based.
Finally, in Section~\ref{sec:expt} we support our results with a case study involving a diffusion process starting from a sparse initial state.
\cut{As an example, one could imagine a situation where a few drops of poison have been introduced into particular (i.e., sparse) locations in a lake of water. From the available measurements at later times, we would like to estimate the source of the contamination.
}

\subsection{Related Work}

Questions involving observability in compressive measurement settings have also been raised in a recent paper~\cite{wangCPF} concerned with tracking the state of a system from nonlinear observations.
Due to the intrinsic nature of the problems in that paper, however, the observability issues raised are quite different.
For example, one argument appears to assume that $M \ge S$, a requirement that we do not have.
In a recent technical report~\cite{dai2012observability}, Dai et al.\ have also considered a similar sparse initial state recovery problem. However, their approach is different and the results are only applicable in noiseless and perfectly sparse initial state recovery problems. In this paper, we establish the \ac{RIP} for the observability matrix, which implies not only that perfectly sparse initial states can be recovered exactly when the measurements are noiseless but also that the recovery process is robust with respect to noise and that nearly-sparse initial states can be recovered with high accuracy~\cite{CandesRIP}.
Finally, we note that a matrix vaguely similar to the observability matrix has been studied by Yap et al.\ in the context of quantifying the memory capacity of echo state networks~\cite{yap2012echo}.
The recovery results and guarantees presented in this paper are substantially different from the above mentioned papers as we derive \ac{CoM} inequalities for the observability matrix and then establish the \ac{RIP} based on these \ac{CoM} inequalities. 
\cut{Moreover, our results guarantee are stable recovery (recovery in presence of noise and for compressible (not perfectly sparse) initial states). 
}
%

\section{Restricted Isometry Property and the Observability Matrix}
\label{sec:rip}

When the observability matrix $\ok$ satisfies the RIP of order $2S$ with isometry constant $\delta_{2S} < \sqrt{2}-1$, an initial state with $S$ or fewer non-zero elements can be stably recovered by solving an $\ell_{1}$-minimization problem~\cite{CandesRIP}. (Similar statements can be made for recovery using various iterative greedy algorithms~\cite{tropp2007signal,needell2009uniform,needell2009cosamp,dai2009subspace}.) In this section, we present cases where the total number of measurements sufficient for establishing the RIP scales linearly in $S$ and only logarithmically in the state dimension $N$.
As in standard observability theory, the state transition matrix $A$ plays a crucial role in the analysis. Because the analysis is somewhat complex, results for completely general $A$ are difficult to obtain. However, we present results here for unitary, scaled unitary, and certain symmetric matrices, and we believe that these can give interesting insight into the essential issues driving the initial state recovery problem.

To assist in interpreting the \ac{RIP} results, let us point out that we actually state our \ac{RIP} bounds in terms of the scaled observability matrix $\frac{1}{\sqrt{b}}\ok$ where $b$ is defined below and is chosen to ensure that
the measurements are properly normalized to be compatible with~(\ref{eq:rip}). In noiseless recovery, this scaling is unimportant. When noise occurs, however, the scaling enters into the effective signal to noise ratio for the problem, as the measurements will be more sensitive to noise when $b$ is small.

Our first result, stated in Theorem~\ref{theo:main_RIP_CoM_result_1}, applies to a system with dynamics represented by a scaled unitary matrix when measurements are taken at the first $K$ sample times, starting at zero.

\begin{theorem}
Assume $\Omega = \left\{0, 1, \dots, K-1\right\}$. Suppose that $A \in \real^{N \times N}$ can be represented as $A = aU$ where $a \in \real$ ($a \neq 0$) and $U\in \real^{N \times N}$ is unitary. Define $b := 1+a^2+a^4+\cdots+a^{2\left(K-1\right)}$. Assume each of the measurement matrices $C_k \in \real^{M \times N}$ is populated with \ac{i.i.d.} Gaussian random entries with mean zero and variance $\frac{1}{M}$. Assume all matrices $C_k$ are generated independently of each other.
Suppose that $N$, $S$, and $\delta_S \in (0,1)$ are given.
Then with probability exceeding $1-\nu$, $\frac{1}{\sqrt{b}}\ok$ satisfies the~\ac{RIP} of order $S$ with isometry constant $\delta_S$ whenever
\begin{numcases}{MK \geq}
\frac{512\left((1-a^2)K+a^2\right)\left(S(\log(\frac{42}{\delta_S})+ 1 + \log(\frac{N}{S})) + \log(\frac{2}{\nu})\right)}{\delta_S^2}, & $|a|<1$ \label{eq:res1a}\\
\frac{512\left((1-a^{-2})K+a^{-2}\right)\left(S(\log(\frac{42}{\delta_S})+ 1 + \log(\frac{N}{S})) + \log(\frac{2}{\nu})\right)}{\delta_S^2}, & $|a|>1$. \label{eq:res1b}
\end{numcases}
\label{theo:main_RIP_CoM_result_1}
\end{theorem}
{\textbf{Proof}} See Section~\ref{sec:RIP_based_CoM}. \hfill $\blacksquare$

One should note that when $A = aU$ ($a \neq 1$), the
results of Theorem~\ref{theo:main_RIP_CoM_result_1} have a dependency on $K$ (the number of sampling times).
This dependency is not desired in general. When $a = 1$ (i.e., $A$ is unitary), a result (Theorem~\ref{theo:main_RIP_CoM_result_3}) can be obtained in which the total number of measurements $MK$ scales linearly in $S$ and with no dependency on $K$. Our general results for $A = aU$ also indicate that when $|a|$ is close to the origin (i.e., $|a| \ll 1$), and by symmetry when $|a| \gg 1$, worse recovery performance is expected as compared to the case when $a = 1$. When $|a| \ll 1$, as an example, the effect of the initial state will be highly attenuated as we take measurements at later times. A similar intuition holds when $|a|\gg1$.
When $a=1$ (i.e., unitary $A$), we can relax the consecutive sample times assumption in Theorem~\ref{theo:main_RIP_CoM_result_1} (i.e., $\Omega = \left\{0, 1, \dots, K-1\right\}$). We have the following \ac{RIP} result when $K$ arbitrarily-chosen samples are taken.

\begin{theorem}
Assume $\Omega = \left\{k_0, k_1, \dots, k_{K-1}\right\}$. Suppose that $A \in \real^{N \times N}$ is unitary.
Assume each of the measurement matrices $C_k \in \real^{M \times N}$ is populated with \ac{i.i.d.} Gaussian random entries with mean zero and variance $\frac{1}{M}$. Assume all matrices $C_k$ are generated independently of each other.
Suppose that $N$, $S$, and $\delta_S \in (0,1)$ are given.
Then with probability exceeding $1-\nu$, $\frac{1}{\sqrt{K}}\ok$ satisfies the~\ac{RIP} of order $S$ with isometry constant $\delta_S$ whenever
\begin{equation}
MK \geq \frac{512\left(S(\log(\frac{42}{\delta_S})+ 1 + \log(\frac{N}{S})) + \log(\frac{2}{\nu})\right)}{\delta_S^2}.\label{eq:res3}
\end{equation}
\label{theo:main_RIP_CoM_result_3}
\end{theorem}
{\textbf{Proof}} See Section~\ref{sec:RIP_based_CoM}. \hfill $\blacksquare$

Theorem~\ref{theo:main_RIP_CoM_result_3} states that under the assumed conditions, $\frac{1}{\sqrt{K}}\ok$ satisfies the \ac{RIP} of order $S$ with isometry constant $\delta_S$ with high probability when the total number of measurements $MK$ scales linearly in the sparsity level $S$ and logarithmically in the state ambient dimension $N$.
Consequently under these assumptions, \emph{unique} recovery of any $S$-sparse initial state $\vc{x}_0$ is possible from $\vc{y}_{\Omega} = \ok\vc{x}_0$ by solving the $\ell_1$-minimization problem or using various iterative greedy algorithms~\cite{tropp2007signal,needell2009cosamp} whenever $MK$ is proportional to $S\log(\frac{N}{S})$. This is in fact a significant reduction in the sufficient total number of measurements for correct initial state recovery as compared to traditional observability theory.

We further extend our analysis and establish the \ac{RIP} for certain symmetric matrices $A$. We believe this analysis has important consequences in analyzing problems of practical interest such as 
diffusion (see, for example, Section~\ref{sec:expt}). Suppose $A \in \real^{N \times N}$ is a positive semidefinite matrix with the eigendecomposition
\begin{equation}
A = U\Lambda U^T=\left[U_1 \big{|} U_2\right]\left[\begin{array}{cc} \Lambda_{1} &0\\0& \Lambda_{2}\end{array} \right]\left[U_1 \big{|} U_2\right]^T,
\label{eq:decomp1}
\end{equation}
where $U \in \real^{N \times N}$ is unitary, $\Lambda \in \real^{N \times N}$ is a diagonal matrix with non-negative entries, $U_1 \in \real^{N \times L}$, $U_2 \in \real^{N \times \left(N-L\right)}$, $\Lambda_1 \in \real^{L \times L}$, and $\Lambda_2 \in \real^{\left(N-L\right) \times \left(N-L\right)}$.
The submatrix $\Lambda_1$ contains the $L$ largest eigenvalues of $A$.
The value for $L$ can be chosen as desired; our results below give the strongest bounds when all eigenvalues in $\Lambda_1$ are large compared to all eigenvalues in $\Lambda_2$.
Let $\lambda_{1,\text{min}}$ denote the smallest entry of $\Lambda_1$, $\lambda_{1,\text{max}}$ denote the largest entry of $\Lambda_1$,
and $\lambda_{2,\text{max}}$ denote the largest entry of $\Lambda_2$.

In the following, we show that in the special case where the matrix $U_1^T \in \real^{L \times N}$ ($L < N$) happens to itself satisfy the \ac{RIP} (up to a scaling), then $\ok$ satisfies the \ac{RIP} (up to a scaling). Although there are many state transition matrices $A$ that do not have a collection of eigenvectors $U_1$ with this special property, we do note that if $A$ is a circulant matrix, its eigenvectors will be the \ac{DFT} basis vectors, and it is known that a randomly selected set of \ac{DFT} basis vectors will satisfy the \ac{RIP} with high probability~\cite{rudelson2008sparse}.

\begin{theorem}
Assume $\Omega = \left\{k_0, k_1, \dots, k_{K-1}\right\}$.
Assume $A$ has the eigendecomposition given in~(\ref{eq:decomp1}) and $U_1^T \in \real^{L \times N}$ ($L < N$) satisfies a scaled version\footnote{The $\frac{L}{N}$
scaling in~(\ref{eq:rip_U_1^T}) is to account for the unit-norm rows of $U_1^T$.} of the \ac{RIP} of order $S$ with isometry constant $\delta_S$. Formally, assume for $\delta_S \in (0,1)$ that
\begin{equation}
(1-\delta_S)\frac{L}{N}\|\vc{x}_0\|_2^2 \leq \|U_1^T\vc{x}_0\|_2^2 \leq (1+\delta_S)\frac{L}{N}\|\vc{x}_0\|_2^2
\label{eq:rip_U_1^T}
\end{equation}
holds for all $S$-sparse $\vc{x}_0 \in \real^N$.
Assume each of the measurement matrices $C_k \in \real^{M \times N}$ is populated with \ac{i.i.d.} Gaussian random entries with mean zero and variance $\frac{1}{M}$. Assume all matrices $C_k$ are generated independently of each other.
Let $\nu \in (0,1)$ denote a failure probability and $\delta \in (0,\frac{16}{\sqrt{K}})$ denote a distortion factor. Then with probability exceeding $1-\nu$,
\begin{equation}
(1-\delta)\left((1-\delta_S)\frac{L}{N}\sum_{i=0}^{K-1}\lambda_{1,\text{min}}^{2k_i}\right) \leq \frac{\|\ok\vc{x}_0\|_2^2}{\|\vc{x}_0\|_2^2} \leq (1+\delta)\left((1+\delta_S)\frac{L}{N}\sum_{i=0}^{K-1}\lambda_{1,\text{max}}^{2k_i} +  \sum_{i=0}^{K-1}\lambda_{2,\text{max}}^{2k_i}\right)
\label{eq:rip_ok_symmetric_A}
\end{equation}
for all $S$-sparse $\vc{x}_0 \in \real^N$ whenever
\begin{equation}
MK \geq \frac{512K\left(S\left(\log(\frac{42}{\delta})+ 1 + \log(\frac{N}{S})\right) + \log(\frac{2}{\nu})\right)}{\rho\delta^2},
\end{equation}
where
\begin{equation}
\rho := \inf_{S-\text{sparse}~\vc{x}_0 \in \real^N} \Gamma(\ak\vc{x}_0)
\label{eq:def_min_Gamma}
\end{equation}
and
\begin{equation*}
\Gamma\left(\ak \vc{x}_0\right) := \frac{\left( \|A^{k_0}\vc{x}_0\|_2^2 + \|A^{k_1} \vc{x}_{0}\|_2^2 + \cdots + \|A^{k_{K-1}} \vc{x}_0\|_2^2\right)^2}{\|A^{k_0}\vc{x}_0\|_2^4 + \|A^{k_1} \vc{x}_{0}\|_2^4 + \cdots + \|A^{k_{K-1}} \vc{x}_0\|_2^4}.
\end{equation*}
\label{theo:rip_ok_symmetric_A}
\end{theorem}
{\textbf{Proof} See Appendix A.} \hfill $\blacksquare$

The result of Theorem~\ref{theo:rip_ok_symmetric_A} is particularly interesting in applications where the largest eigenvalues of $A$ all cluster around each other and the rest of the eigenvalues cluster around zero. Put formally, we are interested in applications where
\[
0 \approx \lambda_{2,\text{max}} \ll \frac{\lambda_{1,\text{min}}}{\lambda_{1,\text{max}}} \approx 1.
\]
The following corollary of Theorem~\ref{theo:rip_ok_symmetric_A} considers an extreme case when $\lambda_{1,\text{max}} = \lambda_{1,\text{min}}$ and $\lambda_{2,\text{max}} = 0$.
\begin{cor}
Assume $\Omega = \left\{k_0, k_1, \dots, k_{K-1}\right\}$.
Assume each of the measurement matrices $C_k \in \real^{M \times N}$ is populated with \ac{i.i.d.} Gaussian random entries with mean zero and variance $\frac{1}{M}$. Assume all matrices $C_k$ are generated independently of each other.
Suppose $A$ has the eigendecomposition given in~(\ref{eq:decomp1}) and $U_1^T \in \real^{L \times N}$ ($L < N$) satisfies a scaled version of the \ac{RIP} of order $S$ with isometry constant $\delta_S$ as given in~(\ref{eq:rip_U_1^T}). Assume $\lambda_{1,\text{max}} = \lambda_{1,\text{min}} =\lambda$ ($\lambda \neq 0$) and $\lambda_{2,\text{max}} = 0$.
Let $\nu \in (0,1)$ denote a failure probability and $\delta \in (0,1)$ denote a distortion factor. Define $C:=\sum_{i=0}^{K-1}\lambda^{2k_i}$ and $\delta^{\prime}_S := \delta_S+\delta+\delta_S\delta$.
Then with probability exceeding $1-\nu$,
\begin{equation}
(1-\delta^{\prime}_S) \|\vc{x}_0\|_2^2 \leq \|\sqrt{\frac{N}{LC}}\ok\vc{x}_0\|_2^2 \leq (1+\delta^{\prime}_S)\|\vc{x}_0\|_2^2
\label{eq:rip_ok_symmetric_A_extreme}
\end{equation}
for all $S$-sparse $\vc{x}_0 \in \real^N$ whenever
\begin{numcases}{MK \geq}
 \frac{512(1+\delta_S)^2\lambda^{-4(k_{K-1}-k_0)}\left(S\left(\log(\frac{42}{\delta})+ 1 + \log(\frac{N}{S})\right) + \log(\frac{2}{\nu})\right)}{(1-\delta_S)^2\delta^2}, & $\lambda<1$\\
 \frac{512(1+\delta_S)^2\lambda^{4(k_{K-1}-k_0)}\left(S\left(\log(\frac{42}{\delta})+ 1 + \log(\frac{N}{S})\right) + \log(\frac{2}{\nu})\right)}{(1-\delta_S)^2\delta^2}, & $\lambda>1$.
\end{numcases}
\label{cor:rip_ok_symmetric_A_extreme}
\end{cor}
{\textbf{Proof}} See Appendix B.\hfill $\blacksquare$

While the result of Corollary~\ref{cor:rip_ok_symmetric_A_extreme} is generic and valid for any $\lambda$, an important \ac{RIP} result can be obtained when $\lambda=1$. The following corollary states the result.

\begin{cor}
Suppose the same notation and assumptions as in Corollary~\ref{cor:rip_ok_symmetric_A_extreme} and additionally assume $\lambda=1$. Then with probability exceeding $1-\nu$,
\begin{equation}
(1-\delta^{\prime}_S) \|\vc{x}_0\|_2^2 \leq \|\sqrt{\frac{N}{LK}}\ok\vc{x}_0\|_2^2 \leq (1+\delta^{\prime}_S)\|\vc{x}_0\|_2^2
\label{eq:rip_ok_symmetric_A_extreme_1}
\end{equation}
for all $S$-sparse $\vc{x}_0 \in \real^N$ whenever
\begin{equation}
MK \geq \frac{512(1+\delta_S)^2\left(S\left(\log(\frac{42}{\delta})+ 1 + \log(\frac{N}{S})\right) + \log(\frac{2}{\nu})\right)}{(1-\delta_S)^2\delta^2}.
\end{equation}
\label{cor:rip_ok_symmetric_A_extreme_1}
\end{cor}
{\textbf{Proof}} See Appendix B.\hfill $\blacksquare$

These results essentially indicate that the more $\lambda$ deviates from one, the more total measurements $MK$ are required to ensure unique recovery of any $S$-sparse initial state $\vc{x}_0$.
The bounds on $\rho$ (which we state in Appendix B to derive Corollaries~\ref{cor:rip_ok_symmetric_A_extreme} and \ref{cor:rip_ok_symmetric_A_extreme_1} from Theorem~\ref{theo:rip_ok_symmetric_A}) also indicate that when $\lambda \neq 1$, the smallest number of measurements are required when the sample times are {\em consecutive} (i.e., when $k_{K-1}-k_{0}=K$).
Similar to what we mentioned earlier in our analysis for a scaled unitary $A$, when $\lambda \neq 1$ the effect of the initial state will be highly attenuated as we take measurements at later times (i.e., when $k_{K-1}-k_0 > K$) which results in a larger total number of measurements $MK$ sufficient for exact recovery.

\section{Concentration of Measure Inequalities and the Observability Matrix}
\label{sec:com}
In this section, we derive \ac{CoM} inequalities for the observability matrix when the measurement matrices $C_k$ are populated with \ac{i.i.d.} Gaussian random entries.
These inequalities are the foundation for establishing the \ac{RIP} presented in the previous section, via Lemma~\ref{lem:RIP_based_CoM}. However, they are also of independent interest for other types of problems involving the states of dynamical systems, such as detection and classification\cite{davenport2010signal,sanandaji2010toeplitz,sanandaji2013com}.
As mentioned earlier, we make a connection to the analysis for block diagonal matrices from~\ac{CS}.
To begin, note that when $\Omega = \left\{k_0, k_1, \dots, k_{K-1}\right\}$ we can write
\begin{equation}
\ok = \underbrace{\left[ \begin{array}{cccc} C_{k_0} &&&\\& C_{k_1} && \\ && \ddots & \\&&& C_{k_{K-1}} \end{array} \right]}_{\ck \in \real^{\km \times \kn}} \underbrace{\left[\begin{array}{c} A^{k_0} \\ A^{k_1} \\ \vdots \\ A^{k_{K-1}} \end{array} \right]}_{\ak \in \real^{\kn \times N}},
\label{eq:ckak}
\end{equation}
where $\widetilde{M} :=MK$ and $\widetilde{N} :=NK$. In this decomposition, $\ck$ is a block diagonal matrix whose diagonal blocks are the measurements matrices, $C_k$. We derive \ac{CoM} inequalities for two cases. We first consider the case where all measurement matrices $C_k$ are generated independently of each other. We then consider the case where all measurement matrices $C_k$ are the same.
\subsection{Independent Random Measurement Matrices}
\label{sec:indep}

In this section, we assume all matrices $C_k$ are generated independently of each other.
Focusing just on $\ck$ for the moment, we have the following bound on its concentration behavior.\footnote{All results in Section~\ref{sec:indep} may be extended to the case where the matrices $C_k$ are populated with sub-Gaussian random variables, as in~\cite{park2011block}.}

\begin{theorem} {\em \cite{park2011block}}
Assume each of the measurement matrices $C_k \in \real^{M \times N}$ is populated with \ac{i.i.d.} Gaussian random entries with mean zero and variance $\frac{1}{M}$. Assume all matrices $C_k$ are generated independently of each other.
Let $\vc{v}_{k_0}, \vc{v}_{k_1}, \dots, \vc{v}_{k_{K-1}} \in \real^N$ and define
$$
\vc{v} = \left[\vc{v}^T_{k_0} ~~ \vc{v}^T_{k_1} ~~ \cdots ~~ \vc{v}^T_{k_{K-1}} \right]^T \in \real^{KN}.
$$
Then
\begin{numcases}{\Prob{\bigg{|}\|\ck \vc{v}\|_{2}^{2}-\|\vc{v}\|_2^2\bigg{|} > \epsilon\|\vc{v}\|_{2}^{2}} \leq}
 2\exp\{-\frac{M\epsilon^{2}\|\vc{\gamma}\|_{1}^{2}}{256\|\vc{\gamma}\|_{2}^{2}}\}, &  $0\leq\epsilon\leq\frac{16\|\vc{\gamma}\|_{2}^{2}}{\|\vc{\gamma}\|_{\infty}\|\vc{\gamma}\|_{1}}$ \label{eq:CoM_lower_eps}\\
2\exp\{-\frac{M\epsilon\|\vc{\gamma}\|_{1}}{16\|\vc{\gamma}\|_{\infty}}\}, & $\epsilon\geq\frac{16\|\vc{\gamma}\|_{2}^{2}}{\|\vc{\gamma}\|_{\infty}\|\vc{\gamma}\|_{1}}$,
\label{eq:CoM_higher_eps}
\end{numcases}
where
$$
\vc{\gamma} = \vc{\gamma}\left(\vc{v}\right) := \left[\begin{array}{c} \|\vc{v}_{k_0}\|_2^2 \\ \|\vc{v}_{k_1}\|_2^2 \\ \vdots \\
\|\vc{v}_{k_{K-1}}\|_2^2 \end{array} \right] \in \real^{K}.
$$
\label{thm:block1}
\end{theorem}

As we will be frequently concerned with applications where
$\epsilon$ is small, consider the first of the cases given in the right-hand side of the above bound. (It can be shown~\cite{park2011block} that this case always permits any value of $\eps$ between $0$ and $\frac{16}{\sqrt{K}}$.) Define
\begin{equation}
 \Gamma =  \Gamma\left(\vc{v}\right)
:= \frac{\|\vc{\gamma}\left(\vc{v}\right)\|_1^2}{\norm{\vc{\gamma}\left(\vc{v}\right)}^2_2}
= \frac{\left( \|\vc{v}_{k_0}\|_2^2 + \|\vc{v}_{k_1}\|_2^2 + \cdots + \|\vc{v}_{k_{K-1}}\|_2^2\right)^2}{\|\vc{v}_{k_0}\|_2^4 + \|\vc{v}_{k_1}\|_2^4 + \cdots + \|\vc{v}_{k_{K-1}}\|_2^4}
\label{eq:Lambda}
\end{equation}
and note that for any $\vc{v} \in
\real^{KN}$, $ 1 \leq  \Gamma\left(\vc{v}\right) \leq K$. This simply follows from the standard relation that $\|\vc{z}\|_2 \leq \|\vc{z}\|_1 \leq \sqrt{K}\|\vc{z}\|_2$ for all $\vc{z} \in \real^K$.
The case $ \Gamma\left(\vc{v}\right) = K$ is quite favorable because the failure probability will decay exponentially fast in the total number of measurements $MK$.
A simple comparison between this result and the \ac{CoM} inequality for a
{\em dense} Gaussian matrix stated in Definition~\ref{def:concgauss} reveals that we get the same degree of concentration from the $MK \times NK$ block diagonal matrix $\ck$ as we would get from a dense $MK \times NK$ matrix populated with \ac{i.i.d.} Gaussian random variables. This event happens if and only if the components $\vc{v}_{k_i}$ have equal energy, i.e., if and only if
$$
\|\vc{v}_{k_0}\|_2 = \|\vc{v}_{k_1}\|_2 = \cdots = \|\vc{v}_{k_{K-1}}\|_2.
$$
On the other hand, the case  $ \Gamma\left(\vc{v}\right) = 1$ is quite unfavorable and implies that we get the same degree of concentration from the $MK \times NK$ block diagonal matrix $\ck$ as we would get from a dense Gaussian matrix having size only $M \times NK$. This event happens if and only if $\|\vc{v}_{k_i}\|_2 = 0$ for all $i \in \left\{0, 1, \dots, K-1\right\}$ but one $i$. Thus, more uniformity in the values of the $\|\vc{v}_{k_i}\|_2$ ensures a higher probability of concentration.

We now note that, when applying the observability matrix to an initial state, we will have
$$
\ok \vc{x}_0 = \ck \ak \vc{x}_0.
$$
This leads us to the following corollary of Theorem~\ref{thm:block1}.
\begin{cor} Suppose the same notation and assumptions as in Theorem~\ref{thm:block1}. Then for any fixed initial state $\vc{x}_0 \in \real^N$ and for any $\eps \in (0,\frac{16}{\sqrt{K}})$,
\begin{eqnarray}
\Prob{\bigg{|}\|\ok \vc{x}_0 \|_{2}^{2}-\|\ak \vc{x}_0\|_2^2\bigg{|} > \epsilon \|\ak \vc{x}_0\|_{2}^{2}}
\leq
2\exp\left\{-\frac{M \Gamma\left(\ak \vc{x}_0\right) \epsilon^{2}}{256}\right\}.
\label{eq:conc2}
\end{eqnarray}
\label{cor:obs_indep}
\end{cor}

There are two important phenomena to consider in this result, and both are impacted by the interaction of $A$ with $\vc{x}_0$.
First, on the left-hand side of (\ref{eq:conc2}), we see that the point of concentration of $\|\ok \vc{x}_0 \|_{2}^{2}$ is around $\|\ak \vc{x}_0\|_2^2$, where
\begin{equation}
\|\ak \vc{x}_0\|_2^2 =\|A^{k_0}\vc{x}_0\|_2^2 + \|A^{k_1} \vc{x}_{0}\|_2^2 + \cdots + \|A^{k_{K-1}} \vc{x}_0\|_2^2.
\label{eq:ak_x_0_sum}
\end{equation}
For a concentration bound of the same form as Definition~\ref{def:concgauss}, however, $\|\ok \vc{x}_0 \|_{2}^{2}$ should concentrate around some constant multiple of $\|\vc{x}_0\|_2^2$. In general, for different initial states $\vc{x}_0$ and transition matrices $A$, we may see widely varying ratios $\frac{\|\ak \vc{x}_0\|_2^2}{\|\vc{x}_0\|_2^2}$. However, further analysis is possible in scenarios where this ratio is predictable and fixed.
Second, on the right-hand side of (\ref{eq:conc2}), we see that the exponent of the concentration failure probability scales with
\begin{equation}
\Gamma\left(\ak \vc{x}_0\right) = \frac{\left( \|A^{k_0}\vc{x}_0\|_2^2 + \|A^{k_1} \vc{x}_{0}\|_2^2 + \cdots + \|A^{k_{K-1}} \vc{x}_0\|_2^2\right)^2}{\|A^{k_0}\vc{x}_0\|_2^4 + \|A^{k_1} \vc{x}_{0}\|_2^4 + \cdots + \|A^{k_{K-1}} \vc{x}_0\|_2^4}.
\label{eq:gamma1}
\end{equation}
As mentioned earlier, $1 \leq  \Gamma\left(\ak \vc{x}_0\right) \leq K$. The case $ \Gamma\left(\ak \vc{x}_0\right) = K$ is quite favorable and happens when $\|A^{k_0}\vc{x}_0\|_2 = \|A^{k_1} \vc{x}_0 \|_2 = \cdots = \|A^{k_{K-1}} \vc{x}_0 \|_2$; this occurs when the state ``energy'' is preserved over time. The case $ \Gamma\left(\ak \vc{x}_0\right) = 1$ is quite unfavorable and happens when $k_0 = 0$ and $\vc{x}_0 \in \mathrm{null}(A)$ for $\vc{x}_0 \neq 0$.


\subsubsection{Unitary and Scaled Unitary System Matrices}
\label{sec:unitary1}

In the special case where $A$ is unitary (i.e., $\| A^{k_i} \vc{x} \|_2^2 = \|\vc{x}\|_2^2$ for all $\vc{x} \in \real^N$ and for any power $k_i$), we can draw a particularly strong conclusion. Because a unitary $A$ guarantees both that $\|\ak \vc{x}_0\|_2^2  = K\|\vc{x}_0\|_2^2$ and that $ \Gamma\left(\ak \vc{x}_0\right) = K$, we have the following result.

\begin{cor} Suppose the same notation and assumptions as in Theorem~\ref{thm:block1}. Assume $\Omega = \left\{k_0, k_1, \dots, k_{K-1}\right\}$.
Suppose that $A$ is a unitary operator. Then for any fixed initial state $\vc{x}_0 \in \real^N$ and for any $\eps \in (0,1)$,\footnote{The observant reader may note that Corollary~\ref{cor:obs_indep} requires $\epsilon$ to be less than $\frac{16}{\sqrt{K}}$. This restriction on $\epsilon$ appears so that we can focus on the upper \ac{CoM} inequality~(\ref{eq:CoM_lower_eps}) and ignore the lower one~(\ref{eq:CoM_higher_eps}). 
However, for most of the problems considered in this paper (i.e., unitary, scaled unitary, and certain symmetric matrices $A$), we can actually apply~(\ref{eq:CoM_lower_eps}) for a much broader range of $\epsilon$ (up 
to $1$ and even higher). In fact, we can show that in these settings,
\[
2\exp\{-\frac{M\epsilon^{2}\|\vc{\gamma}\|_{1}^{2}}{256\|\vc{\gamma}\|_{2}^{2}}\} \geq 2\exp\{-\frac{M\epsilon\|\vc{\gamma}\|_{1}}{16\|\vc{\gamma}\|_{\infty}}\}
\]
for all $\epsilon \in (0,1)$. Consequently, we allow $\epsilon \in (0,1)$ in Corollaries~\ref{cor:unitary1} and \ref{cor:scaled-unitary1}. We have omitted these details for the sake of space.} 
\begin{equation}
\Prob{\bigg{|}\|\frac{1}{\sqrt{K}}\ok \vc{x}_0 \|_{2}^{2}- \|\vc{x}_0\|_2^2\bigg{|} > \epsilon  \|\vc{x}_0\|_{2}^{2}}
\leq 2\exp\left\{-\frac{M K \epsilon^{2}}{256}\right\}.
\label{eq:conc3}
\end{equation}
\label{cor:unitary1}
\end{cor}

What this means is that we get the same degree of concentration from the $MK \times N$ matrix $\frac{1}{\sqrt{K}}\ok$ as we would get from a
fully dense $MK \times N$ matrix populated with \ac{i.i.d.} Gaussian random variables.
Observe that this concentration result is valid for any
$\vc{x}_0 \in \real^N$ (not necessarily sparse) and can be used, for example, to prove that finite point clouds~\cite{indyk1998approximate} and low-dimensional manifolds~\cite{mbwFocm} in $\real^N$ can have stable, approximate distance-preserving embeddings under the matrix $\frac{1}{\sqrt{K}}\ok$. In each of these cases we may be able to solve very powerful signal inference and recovery problems with $MK \ll N$.

When $\Omega = \left\{0, 1, \dots, K-1\right\}$ (consecutive sample times), one can further derive \ac{CoM} inequalities when $A$ is a scaled unitary matrix (i.e., when $A = aU$ where $a \in \real$ ($a \neq 1$) and $U \in \real^{N \times N}$ is unitary).

\begin{cor} Suppose the same notation and assumptions as in Theorem~\ref{thm:block1}. Assume $\Omega = \left\{0, 1, \dots, K-1\right\}$. Suppose that $A = aU$ ($a \in \real, a \neq 0$) and $U \in \real^{N \times N}$ is unitary. Define $b:=\Sigma_{k=0}^{K-1}a^{2k}$.
Then for any fixed initial state $\vc{x}_0 \in \real^N$ and for any $\eps \in (0,1)$,
\begin{numcases}{\Prob{\bigg{|}\|\frac{1}{\sqrt{b}}\ok \vc{x}_0 \|_{2}^{2}- \|\vc{x}_0\|_2^2\bigg{|} > \epsilon  \|\vc{x}_0\|_{2}^{2}}
\leq}
2\exp\left\{-\frac{M K \epsilon^{2}}{256\left(\left(1-a^2\right)K+a^2\right)}\right\}, & $|a|<1$ \label{eq:scaled-conc3}\\
2\exp\left\{-\frac{M K \epsilon^{2}}{256\left(\left(1-a^{-2}\right)K+a^{-2}\right)}\right\}, & $|a|>1$. \label{eq:scaled-conc4}
\end{numcases}
\label{cor:scaled-unitary1}
\end{cor}

{\textbf{Proof of Corollary \ref{cor:scaled-unitary1}}}
First note that when $A=aU$ ($U$ is unitary) and $\Omega = \left\{0, 1, \dots, K-1\right\}$ then $\|\ak\vc{x}_0\|_2^2 = (1+a^2+\cdots+a^{2\left(K-1\right)})\|\vc{x}_0\|_2^2 = b\|\vc{x}_0\|_2^2$. From~(\ref{eq:gamma1}) when $|a|<1$,
\begin{equation}
\Gamma\left(\ak \vc{x}_0\right) = \frac{\left( 1 + a^2 + \cdots + a^{2\left(K-1\right)}\right)^2}{1 + a^4 + \cdots + a^{4\left(K-1\right)}} = \frac{\left(1-a^{2K}\right)\left(1+a^2\right)}{\left(1+a^{2K}\right)\left(1-a^2\right)} = \frac{\frac{1+a^2}{1+a^{2K}}}{\frac{1-a^2}{1-a^{2K}}}.
\label{eq:bnd_Gamma_a_small_1}
\end{equation}
Also observe\footnote{In order to prove~(\ref{eq:geo_sum_bnd_1}), for a given $|a|<1$, let $C\left(a\right)$ be a constant such that for all $K$ ($K$ only takes positive integer values), $\frac{1}{1-a^{2K}} \leq 1+\frac{C\left(a\right)}{K}$. By this assumption, $C\left(a\right) \geq \frac{Ka^{2K}}{1-a^{2K}} =: g\left(a,K\right)$.
Observe that for a given $|a|<1$, $g\left(a,K\right)$ is a decreasing function of $K$ and its maximum is achieved when $K=1$. Choosing $C\left(a\right) = g\left(a,1\right) = \frac{a^2}{1-a^2}$ completes the proof of~(\ref{eq:geo_sum_bnd_1}).}
that when $|a|<1$,
\begin{equation}
\frac{1-a^2}{1-a^{2K}} \leq (1-a^2)+\frac{a^2}{K}.
\label{eq:geo_sum_bnd_1}
\end{equation}
Thus, from~(\ref{eq:bnd_Gamma_a_small_1}) and (\ref{eq:geo_sum_bnd_1}) and noting that $1+a^2 \geq 1+a^{2K}$ when $|a|<1$,
\[
\Gamma\left(\ak \vc{x}_0\right) \geq \frac{K}{(1-a^2)K+a^2}.
\]
Similarly, one can show that when $|a|>1$,
\begin{equation}
\frac{1-a^{-2}}{1-a^{-2K}} \leq (1-a^{-2})+\frac{a^{-2}}{K}
\label{eq:geo_sum_bnd_2}
\end{equation}
and consequently,
\[
\Gamma\left(\ak \vc{x}_0\right) \geq \frac{K}{(1-a^{-2})K+a^{-2}}.
\]
With the appropriate scaling of $\ok$ by $\frac{1}{\sqrt{b}}$, the \ac{CoM} inequalities follow from Corollary~\ref{cor:obs_indep}.
\hfill $\blacksquare$

%

\subsubsection{Implications for the \ac{RIP}}
\label{sec:RIP_based_CoM}
As mentioned earlier, our \ac{CoM} inequalities have immediate implications in establishing the \ac{RIP} for the observability matrix. Based on Definition~\ref{def:concgauss} and Lemma~\ref{lem:RIP_based_CoM}, in this section we prove Theorems~\ref{theo:main_RIP_CoM_result_1}
and \ref{theo:main_RIP_CoM_result_3}.

{\textbf{Proof of Theorem~\ref{theo:main_RIP_CoM_result_1}}} In order to establish the \ac{RIP} based on Lemma~\ref{lem:RIP_based_CoM}, we simply need to evaluate $f(\epsilon)$ in our \ac{CoM} result derived in Corollary~\ref{cor:scaled-unitary1}. One can easily verify that
\begin{numcases}{f(\epsilon) = }
\frac{\epsilon^2}{256\left((1-a^2)K+a^2\right)}, & $|a|<1$\\
\frac{\epsilon^2}{256\left((1-a^{-2})K+a^{-2}\right)}, & $|a|>1$.
\end{numcases}
Through a union bound argument and by applying Lemma~\ref{lem:RIP_based_CoM} for all $N \choose S$ $S$-dimensional subspaces in $\real^N$, the \ac{RIP} result follows.
\hfill $\blacksquare$

{\textbf{Proof of Theorem~\ref{theo:main_RIP_CoM_result_3}}} In order to establish the \ac{RIP} based on Lemma~\ref{lem:RIP_based_CoM}, we simply need to evaluate $f(\epsilon)$ in our \ac{CoM} result derived in Corollary~\ref{cor:unitary1}. In this case,
\[
f(\epsilon) = \frac{\epsilon^2}{256}.
\]
Through a union bound argument and by applying Lemma~\ref{lem:RIP_based_CoM} for all $N \choose S$ $S$-dimensional subspaces in $\real^N$, the \ac{RIP} result follows.
\hfill $\blacksquare$

\subsection{Identical Random Measurement Matrices}
\label{sec:same}

In this section, we consider the case where all matrices $C_k$ are identical and equal to some $M \times N$ matrix $C$ which is populated with \ac{i.i.d.} Gaussian entries having zero mean and variance $\sigma^2 = \frac{1}{M}$. Once again note that we can write
$
\ok = \ck\ak,
$
where this time
\begin{equation} \small
\ck:=  \left[ \begin{array}{cccc} C_{k_0} &&&\\& C_{k_1} && \\ && \ddots & \\&&& C_{k_{K-1}} \end{array} \right] = \left[ \begin{array}{cccc} C &&&\\& C && \\ && \ddots & \\&&& C \end{array} \right],
\label{eq:ck2}
\end{equation}
and $\ak$ is as defined in~(\ref{eq:ckak}). The matrix $\ck$ is block diagonal with equal blocks on its main diagonal, and we have the following bound on its concentration behavior.

\begin{theorem} {\em \cite{park2011block}} Assume each of the measurement matrices $C_k \in \real^{M \times N}$ is populated with \ac{i.i.d.} Gaussian random entries with mean zero and variance $\frac{1}{M}$. Assume all matrices $C_k$ are the same (i.e., $C_k = C, \forall k$). Let $\vc{v}_{k_0}, \vc{v}_{k_1}, \dots, \vc{v}_{k_{K-1}} \in \real^N$ and define
$$
\vc{v} = \left[\vc{v}^T_{k_0} ~~ \vc{v}^T_{k_1} ~~ \cdots ~~ \vc{v}^T_{k_{K-1}} \right]^T \in \real^{KN}.
$$
Then,
\begin{equation}
\Prob{\bigg{|}\|\ck \vc{v}\|_{2}^{2}-\|\vc{v}\|_2^2\bigg{|} > \epsilon\|\vc{v}\|_{2}^{2}}
\leq
\begin{cases}
2\exp\{-\frac{M\epsilon^{2}\|\vc{\lambda}\|_{1}^{2}}{256\|\vc{\lambda}\|_{2}^{2}}\},& 0\leq\epsilon\leq\frac{16\|\vc{\lambda}\|_{2}^{2}}{\|\vc{\lambda}\|_{\infty}\|\vc{\lambda}\|_{1}} \\
2\exp\{-\frac{M\epsilon\|\vc{\lambda}\|_{1}}{16\|\vc{\lambda}\|_{\infty}}\},&
\epsilon\geq\frac{16\|\vc{\lambda}\|_{2}^{2}}{\|\vc{\lambda}\|_{\infty}\|\vc{\lambda}\|_{1}},
\end{cases}
\end{equation}
where
$$
\vc{\lambda} = \vc{\lambda}\left(\vc{v}\right) := \left[\begin{array}{c} \lambda_1 \\ \lambda_2 \\ \vdots \\ \lambda_{\min(K,N)}
\end{array} \right] \in \real^{{\min(K,N)}},
$$
and $\{\lambda_1, \lambda_2, \dots, \lambda_{\min(K,N)}\}$ are the first (non-zero) eigenvalues of the $K \times K$ matrix $V^T V$, where
\[
V = \left[\vc{v}_{k_0} ~~ \vc{v}_{k_1} ~~ \cdots ~~ \vc{v}_{k_{K-1}} \right]\in \real^{N \times K}.
\]
\label{thm:block2}
\end{theorem}

Consider the first of the cases given in the right-hand side of the above bound. (This case permits any value of $\epsilon$ between $0$ and $\frac{16}{\sqrt{\min(K,N)}}$.) Define
\begin{equation}
\Lambda\left(\vc{v}\right) := \frac{\|
\vc{\lambda}\left(\vc{v}\right)\|_1^2}{\norm{\vc{\lambda}\left(\vc{v}\right)}^2_2}
\label{eq:Lambda2}
\end{equation}
and note that for any $\vc{v} \in
\real^{NK}$, $ 1 \le \Lambda\left(\vc{v}\right) \le \min(K,N)$. Moving forward, we will assume for simplicity that $K \le N$, although this assumption can be removed.
The case $\Lambda\left(\vc{v}\right) = K$ is quite favorable and implies that we get the same degree of concentration from the $MK \times NK$
block diagonal matrix $\ck$ as we would get from a dense $MK \times NK$ matrix populated with \ac{i.i.d.} Gaussian random variables. This event happens if and only if $\lambda_1 = \lambda_2 = \cdots = \lambda_K$, which happens if and only if
$$
\|\vc{v}_{k_0}\|_2 = \|\vc{v}_{k_1}\|_2 = \cdots = \|\vc{v}_{k_{K-1}}\|_2
$$
and $\langle \vc{v}_{k_i}, \vc{v}_{k_\ell} \rangle = 0$ for all $0 \le i,\ell \le K-1$ with $i \neq \ell$.
On the other hand, the case  $\Lambda\left(\vc{v}\right) = 1$ is quite unfavorable and implies that we get the same degree of concentration from the $MK \times NK$ block diagonal matrix $\ck$ as we would get from a dense Gaussian matrix having only $M$ rows.
This event happens if and only if the dimension of $\mathrm{span}\{\vc{v}_{k_0}, \vc{v}_{k_1}, \dots, \vc{v}_{k_{K-1}} \}$ equals 1. Thus, comparing to Section~\ref{sec:indep}, uniformity in the norms of the vectors $\vc{v}_k$ is no longer sufficient for a high probability of concentration; in addition to this we must have diversity in the directions of the $\vc{v}_{k_i}$.

The following corollary of Theorem~\ref{thm:block2} derives
a \ac{CoM} inequality for the observability matrix. Recall that $\ok \vc{x}_0 = \ck \ak \vc{x}_0$ where $\ck$ is a block diagonal matrix whose diagonal blocks are repeated.

\begin{cor} Suppose the same notation and assumptions as in Theorem~\ref{thm:block2} and suppose $K \le N$. Then for any fixed initial state $\vc{x}_0 \in \real^N$ and for any $\eps \in (0,\frac{16}{\sqrt{K}})$,
\begin{equation}
\Prob{\bigg{|}\|\ok \vc{x}_0 \|_{2}^{2}-\|\ak \vc{x}_0\|_2^2\bigg{|}> \epsilon\|\ak \vc{x}_0\|_{2}^{2}}
\leq
2\exp\left\{-\frac{M \Lambda(\ak \vc{x}_0) \epsilon^{2}}{256}\right\}.
\label{eq:conc4}
\end{equation}
\end{cor}

Once again, there are two important phenomena to consider in this result, and both are impacted by the interaction of $A$ with $\vc{x}_0$.
First, on the left hand side of (\ref{eq:conc4}), we see that the point of concentration of $\|\ok \vc{x}_0 \|_{2}^{2}$ is around $\|\ak \vc{x}_0\|_2^2$. Second, on the right-hand side of (\ref{eq:conc4}), we see that the exponent of the concentration failure probability scales with $\Lambda(\ak \vc{x}_0)$, which is determined by the eigenvalues of the $K \times K$ Gram matrix $V^T V$, where
$$
V = \left[A^{k_0}\vc{x}_0 ~~ A^{k_1}\vc{x}_0 ~~ \cdots ~~ A^{k_{K-1}}\vc{x}_0 \right]\in \real^{N \times K}.
$$
As mentioned earlier, $1 \le \Lambda(\ak \vc{x}_0) \le K$. The case $\Lambda(\ak \vc{x}_0) = K$ is quite favorable and happens when $\|A^{k_0}\vc{x}_0\|_2 = \|A^{k_1}\vc{x}_0 \|_2 = \cdots = \|A^{k_{K-1}}\vc{x}_0 \|_2$ and $\langle A^{k_i} \vc{x}_0 , A^{k_{\ell}} \vc{x}_0 \rangle = 0$ for all $0 \le i,\ell \le K-1$ with $i \neq \ell$. The case $\Lambda(\ak \vc{x}_0) = 1$ is quite unfavorable and happens if the dimension of $\mathrm{span}\{A^{k_0}\vc{x}_0, A^{k_1}\vc{x}_0 , \dots, A^{k_{K-1}}\vc{x}_0 \}$ equals 1.

In the special case where $A$ is unitary, we know that $\|\ak \vc{x}_0\|_2^2  = K\|\vc{x}_0\|_2^2$. However, a unitary system matrix does not guarantee a favorable value for $\Lambda(\ak \vc{x}_0)$. Indeed, if $A=I_{N \times N}$ we obtain the worst case value $\Lambda(\ak \vc{x}_0) = 1$. If, on the other hand, $A$ acts as a rotation that takes a state into an orthogonal subspace, we will have a stronger result.

\begin{cor} Suppose the same notation and assumptions as in Theorem~\ref{thm:block2} and suppose $K \le N$. Suppose that $A$ is a unitary operator. Suppose also that $\langle A^{k_i} \vc{x}_0 , A^{k_{\ell}} \vc{x}_0 \rangle = 0$ for all $0 \le i,\ell \le K-1$ with $i \neq \ell$. Then for any fixed initial state $\vc{x}_0 \in \real^N$ and for any $\eps \in (0,\frac{16}{\sqrt{K}})$,
\begin{equation}
\Prob{\bigg{|}\|\frac{1}{\sqrt{K}}\ok \vc{x}_0 \|_{2}^{2}- \|\vc{x}_0\|_2^2\bigg{|}> \epsilon  \|\vc{x}_0\|_{2}^{2}}
\leq
2\exp\left\{-\frac{M K \epsilon^{2}}{256}\right\}.
\label{eq:conc5}
\end{equation}
\label{cor:indep_unitary_rotate}
\end{cor}

This result requires a particular relationship between $A$ and $\vc{x}_0$, namely that $\langle A^{k_i} \vc{x}_0 , A^{k_{\ell}} \vc{x}_0 \rangle = 0$ for all $0 \le i,\ell \le K-1$ with $i \neq \ell$. Thus, given a particular system matrix $A$, it is possible that it might hold for some $\vc{x}_0$ and not others. One must therefore be cautious in using this concentration result for CS applications (such as proving the RIP) that involve applying the concentration bound to a prescribed collection of vectors~\cite{baraniuk2008simple}; one must ensure that the ``orthogonal rotation'' property holds for each vector in the prescribed set.
\section{Case Study: Estimating the Initial State in a Diffusion Process}
\label{sec:expt}

So far we have provided theorems that provide a sufficient number of measurements for stable recovery of a sparse initial state under certain conditions on the state transition matrix and under the assumption that the measurement matrices are independent and populated with random entries. In this section, we use a case study to illustrate some of the phenomena raised in the previous sections.
%
\subsection{System Model}

We consider the problem of estimating the initial state of a system governed by the diffusion equation
\[
\frac{\partial x}{\partial t} = \nabla \cdot\left( D(p) \nabla x(p,t)\right),
\]
where $x(p,t)$ is the concentration, or density, at position $p$ at time $t$, and $D(p)$ is the diffusion coefficient at position $p$. If $D$ is independent of position, then this simplifies to
\[
\frac{\partial x}{\partial t} = D \nabla^2 x(p,t).
\]
The boundary conditions can vary according to the surroundings of the domain $\Pi$. If $\Pi$ is bounded by an impermeable surface (e.g., a lake surrounded by the shore), then the boundary conditions are $\left. n(p) \cdot \frac{\partial x}{\partial p} \right|_{p \in\partial \Pi}=0$, where $n(p)$ is the normal to $\partial \Pi$ at $p$.
We will work with an approximate model discretized in time and in space. For simplicity, we explain a one-dimensional (one spatial dimension) diffusion process here but a similar approach of discretization can be taken for a diffusion process with two or three spatial dimensions.
Let $\vc{p}:=\begin{bmatrix} p(1) & p(2) & \cdots & p(N)\end{bmatrix}^T$ be a vector of equally-spaced locations with spacing $\Delta_s$, and let $\vc{x}\left(p,t\right):=\begin{bmatrix} x(p(1),t) & x(p(2),t) & \cdots & x(p(N),t) \end{bmatrix}^T$. Then a first difference approximation in space gives the model
\begin{equation}
\dot{\vc{x}}\left(p,t\right) = G \vc{x}\left(p,t\right),
\label{eqn:conttime}
\end{equation}
where $G$ represents the discrete Laplacian. We have
\[
G = -F = \frac{D}{\Delta_s^2}\begin{bmatrix} -1 & 1 & 0 & 0 & \cdots & 0 \\
1 & -2 & 1 & 0 & \cdots & 0 \\
0 & 1 & -2 & 1 & \cdots & 0 \\
\vdots & &  \ddots & \ddots & \ddots & \vdots \\
0 & \cdots &   & & 1 & -1 \end{bmatrix},
\]
where $F$ is the Laplacian matrix associated with a path (one spatial dimension).
This discrete Laplacian $G$ has eigenvalues $\lambda_{i}=-\frac{2D}{\Delta^{2}_{s}}\left(1 - \cos\left(\frac{\pi}{N}(i-1)\right)\right)$ for $i=1, 2, \dots, N$.

\begin{figure}
\begin{center}
\includegraphics[width=3in]{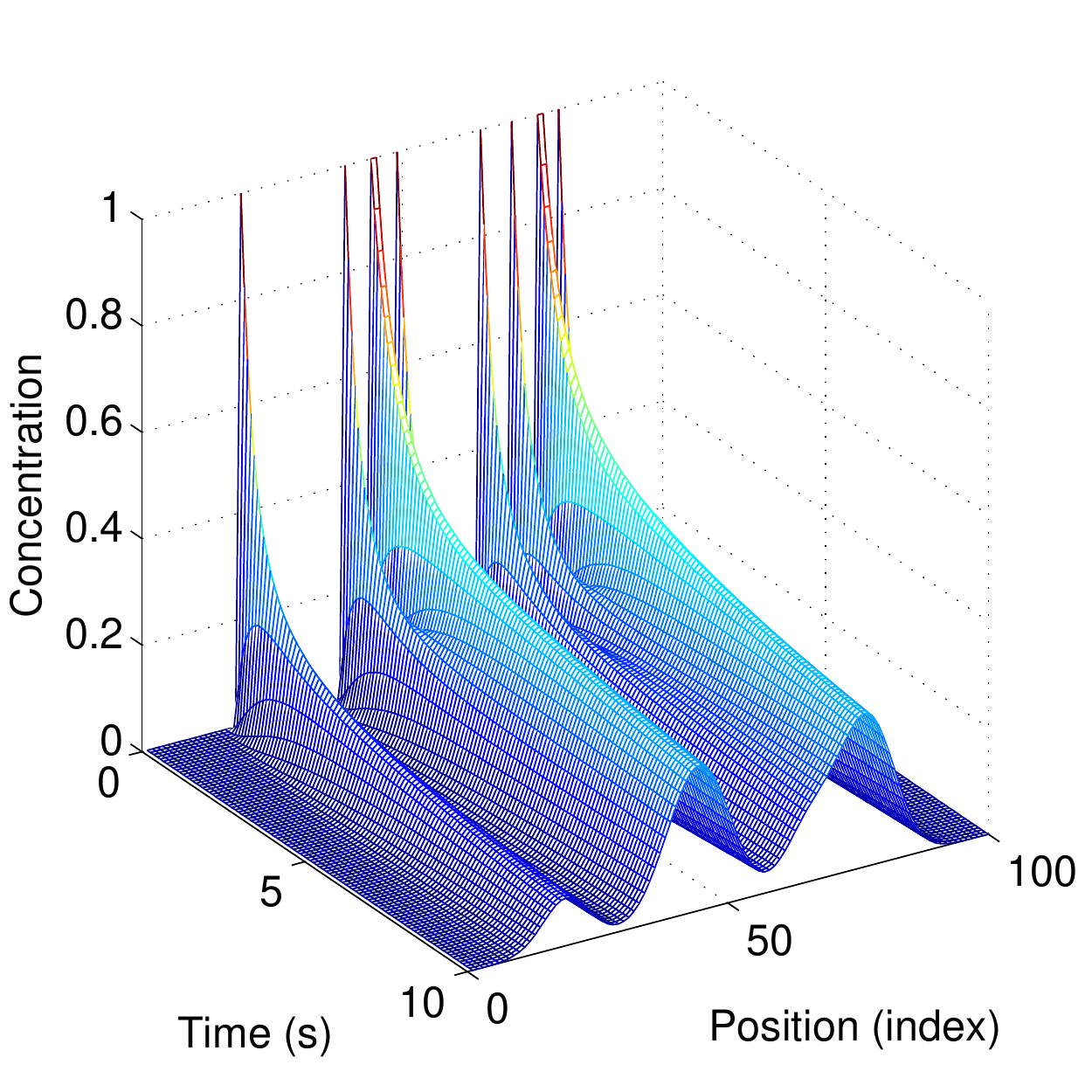}
\caption{One-dimensional diffusion process. At time zero, the concentration (the state) is non-zero only at a few locations of the path graph of $N=100$ nodes.}
\label{fig:diffsim1}
\ss
\end{center}
\end{figure}

To obtain a discrete-time model, we choose a sampling time $T_{s}$ and let the vector
$\vc{x}_{k}=\vc{x}(p,kT_{s})$ be the concentration at positions $p(1), p(2), \dots, p(N)$ at sampling time $k$.  Using a first difference approximation in time, we have
\[
\vc{x}_{k} = A \vc{x}_{k-1},
\label{eqn:disctime}
\]
where $A = I_N+ G T_{s}$.
For a diffusion process with two spatial dimensions, a similar analysis would follow, except one would use the Laplacian matrix of a grid (instead of the Laplacian matrix of a one-dimensional path) in $A = I_N+ G T_{s}$.
For all simulations in this section we take $D=1$, $\Delta_{s}=1$, $N = 100$, and $T_{s} = 0.1$. An example simulation of a one-dimensional diffusion is shown in Figure~\ref{fig:diffsim1}, where we have initialized the system with a sparse initial state $\vc{x}_0$ containing unit impulses at $S = 10$ randomly chosen locations.

In Section~\ref{sec:exprecovery}, we provide several simulations which demonstrate that recovery of a sparse initial state is possible from compressive measurements.

\subsection{Diffusion and its Connections to Theorem~\ref{theo:rip_ok_symmetric_A}}

Before presenting the recovery results from compressive measurements, we would like to mention that our analysis in Theorem~\ref{theo:rip_ok_symmetric_A} gives some insight into (but is not precisely applicable to) the diffusion problem. In particular, the discrete Laplacian matrix $G$ and the corresponding state transition matrix $A$ (see below) are almost circulant, and so their eigenvectors will closely resemble the \ac{DFT} basis vectors. The largest eigenvalues correspond to the lowest frequencies, and so the $U_1$ matrix corresponding to $G$ or $A$ will resemble a basis of the lowest frequency \ac{DFT} vectors. While such a matrix does not technically satisfy the \ac{RIP}, matrices formed from random sets of \ac{DFT} vectors do satisfy the \ac{RIP} with high probability~\cite{rudelson2008sparse}. Thus, even though we cannot apply Theorem~\ref{theo:rip_ok_symmetric_A} directly to the diffusion problem, it does provide some intuition that sparse recovery should be possible in the diffusion setting.


\subsection{State Recovery From Compressive Measurements}
\label{sec:exprecovery}

In this section, we consider a two-dimensional diffusion process. As mentioned earlier, the state transition matrix $A$ associated with this process is of the form $A = I_N + GT_s$, where $T_s$ is the sampling time and $G$ is the Laplacian matrix of a grid. In these simulations, we consider a grid of size $10 \times 10$ with $T_s = 0.1$.

\begin{figure}[tb]
\centering
\subfigure[]{
   \includegraphics[width = .45\columnwidth]{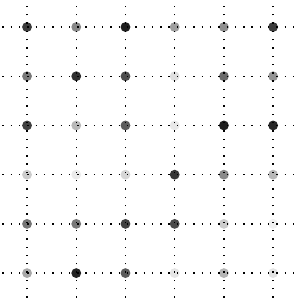}
   \label{fig:Dense_Meas}
 }
\subfigure[]{
   \includegraphics[width = .45\columnwidth]{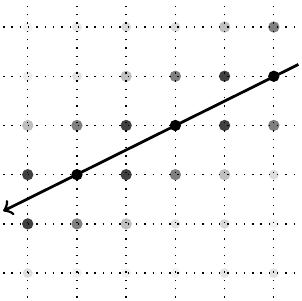}
   \label{fig:Line_Meas}
 }
\caption{
Dense Measurements versus Line Measurements.
The color of a node indicates the corresponding weight of that node. The darker the node color, the higher the weight. These weights are the entries of each row of each $C_k$. (a) Dense Measurements. The weights are drawn from a Gaussian distribution with mean zero and variance $\frac{1}{M}$. These values are random and change for each measurement.
(b) Line Measurements. The weights are generated as a function of the perpendicular distances of all nodes of the grid to the line. The slope and the intercept of the line are random and change for each measurement.}
\label{fig:Dense_Line_Meas}
\end{figure}

We also consider two types of measuring processes. We first look at random measurement matrices $C_k \in \real^{M \times N}$ where the entries of each matrix are \ac{i.i.d.} Gaussian random variables with mean zero and variance $\frac{1}{M}$.
Note that this type of measurement matrix falls within the assumptions of our theorems in Sections~\ref{sec:rip} and \ref{sec:indep}. In this measuring scenario, all of the nodes of the grid (i.e., all of the states) will be measured at each sample time. Formally, at each observation time we record a random linear combination of all nodes.
In the following, we refer to such measurements as ``Dense Measurements.'' Figure~\ref{fig:Dense_Meas} illustrates an example of how the random weights are spread over the grid. The weights (the entries of each row of each $C_k$) are shown using grayscale. The darker the node color, the higher the corresponding weight.
We also consider a more practical measuring process in which at each sample time the operator measures the nodes of the grid occurring along a line with random slope and random intercept. Formally, $C_k\left(i,j\right) = \exp\left(-\frac{d_k\left(i,j\right)}{c}\right)$
where $d_k\left(i,j\right)$ is the perpendicular
distance of node $j$ ($j = 1, \dots, N$) to the $i$th ($i = 1, \dots, M$) line with random slope and random intercept and $c$ is an absolute constant that determines how fast the node weights decrease as their distances increase from the line. Figure~\ref{fig:Line_Meas} illustrates an example of how the weights are spread over the grid in this scenario. Observe that the nodes that are closer to the line are darker, indicating higher weights for those nodes.\cut{
The slope and the intercept of the line is random for each measurement.
Put formally, the entries of each row of each measurement matrix $C_k \in \real^{M \times N}$ are generated as a function of the perpendicular distances of all nodes of the grid to a random line. A random line in this context is a line with a random slope and a random intercept that passes through the field. In other words, in order to take a measurement in this scenario, we first generate a random line and then calculate the perpendicular distances of all the points of the grid to this line. The entries of the corresponding row of $C_k$ are functions of these distances. We use an exponential function for this regard.} We refer to such measurements as ``Line Measurements.''

To address the problem of recovering the initial state $\vc{x}_0$, let us first consider the situation where we collect measurements only of $\vc{x}_0 \in \real^{100}$ itself. We fix the sparsity level of $\vc{x}_0$ to $S = 9$.
For various values of $M$, we construct measurement matrices $C_0$ according to the two models explained above.
At each trial, we collect the measurements $\vc{y}_0 = C_0 \vc{x}_0$ and attempt to recover $\vc{x}_0$ given $\vc{y}_0$ and $C_0$ using the canonical $\ell_1$-minimization problem from CS:
\begin{equation}
\widehat{\vc{x}}_0 = \arg\min_{\vc{x} \in \real^N} \|\vc{x}\|_1 ~~\mathrm{subject~to} ~~ \vc{y}_k = C_k A^{k} \vc{x}
\label{eq:ell1k}
\end{equation}
with $k=0$. (In the next paragraph, we repeat this experiment for different $k$.)
In order to imitate what might happen in reality (e.g., a drop of poison being introduced to a lake of water at $k  = 0$), we assume the initial contaminant appears in a cluster of nodes on the associated diffusion grid. In our simulations, we assume the $S=9$ non-zero entries of the initial state correspond to a $3 \times 3$
square-neighborhood of nodes on the grid. For each $M$, we repeat the recovery problem for $300$ trials; in each trial we generate a random sparse initial state $\vc{x}_0$ (an initial state with a random location of the $3 \times 3$ square and random values of the $9$ non-zero entries) and a measurement matrix $C_0$ as explained above.

\begin{figure}[tb]
\centering
\subfigure[]{
   \includegraphics[width = .45\columnwidth]{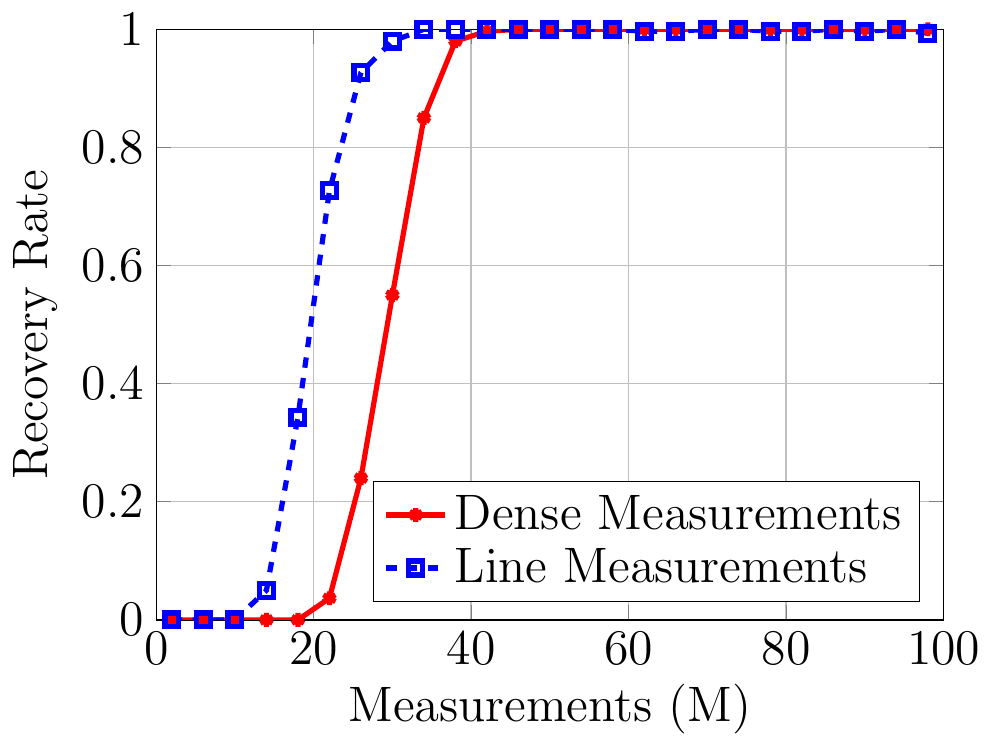}
   \label{fig:NEW_MEAS_TwoDimDiffRecoveryTest2_Tvec_0}
 }
\subfigure[]{
   \includegraphics[width = .45\columnwidth]{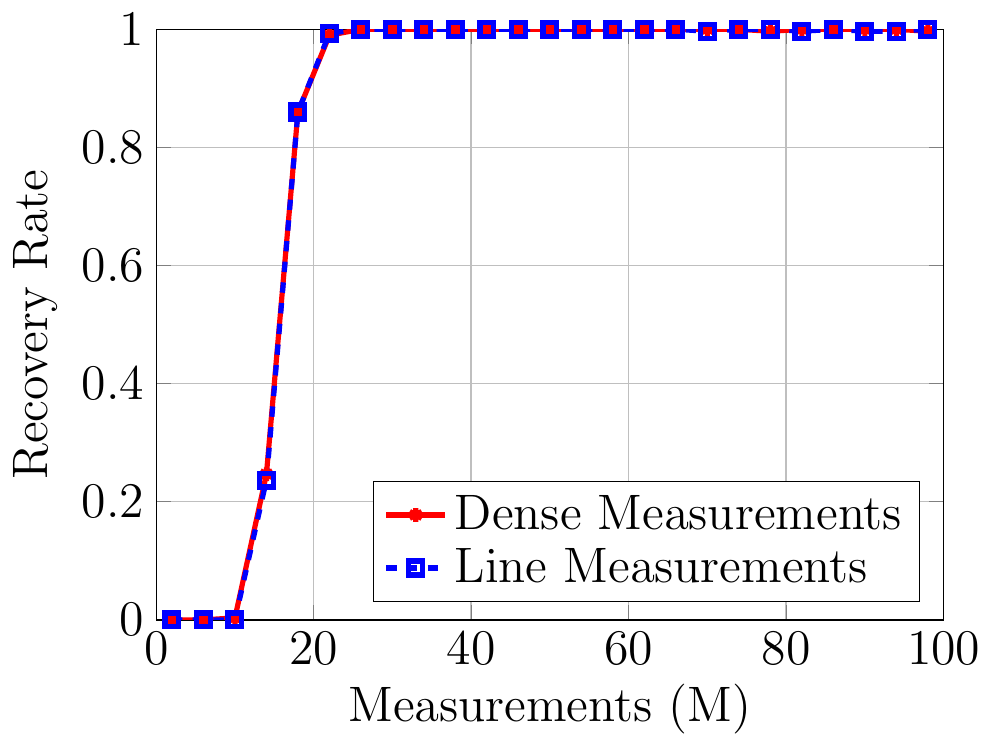}
   \label{fig:NEW_MEAS_TwoDimDiffRecoveryTest2_Tvec_10}
 }
\caption{
Signal recovery from compressive measurements of a diffusion process which has initiated from a sparse initial state of dimension $N = 100$ and sparsity level $S = 9$. The plots show the percent of trials (out of $300$ trials in total) with perfect recovery of the initial state $\vc{x}_0$ versus the number of measurements $M$. (a) Recovery from compressive measurements at time $k=0$. (b) Recovery from compressive measurements at time $k=10$.
}
\label{fig:TwoDimDiffRecoveryTest2_Tvec_0_1}
\end{figure}

Figure~\ref{fig:NEW_MEAS_TwoDimDiffRecoveryTest2_Tvec_0} depicts, as a function of $M$, the percent of trials (with $\vc{x}_0$ and $C_0$ randomly chosen in each trial) in which the initial state is recovered perfectly, i.e., $\widehat{\vc{x}}_0 = \vc{x}_0$. Naturally, we see that as we take more measurements, the recovery rate increases.
When Line Measurements are taken, with almost $35$ measurements we recover every sparse initial state of dimension $100$ with sparsity level $9$. When Dense Measurements are employed, however, we observe a slightly weaker recovery performance at $k = 0$ as almost $45$ measurements are required to see exact recovery.
In order to see how the diffusion phenomenon affects the recovery, we repeat the same experiment at $k=10$. In other words, we collect the measurements $\vc{y}_{10} = C_{10}\vc{x}_{10} = C_{10} A^{10}\vc{x}_0$ and attempt to recover $\vc{x}_0$ given $\vc{y}_{10}$ and $C_{10}A^{10}$ using the canonical $\ell_1$-minimization problem~(\ref{eq:ell1k}). As shown in Fig.~\ref{fig:NEW_MEAS_TwoDimDiffRecoveryTest2_Tvec_10}, the recovery performance is improved when Line and Dense Measurements are employed (with almost $25$ measurements exact recovery is possible).
Qualitatively, this suggests that due to diffusion, at $k=10$, the initial contaminant is now propagating and consequently a larger surface of the lake (corresponding to more nodes of the grid) is contaminated. In this situation, a higher number of contaminated nodes will be measured by Line Measurements which potentially can improve the recovery performance of the initial state.

\begin{figure}[tb]
\centering
\subfigure[]{
   \includegraphics[width = .45\columnwidth]{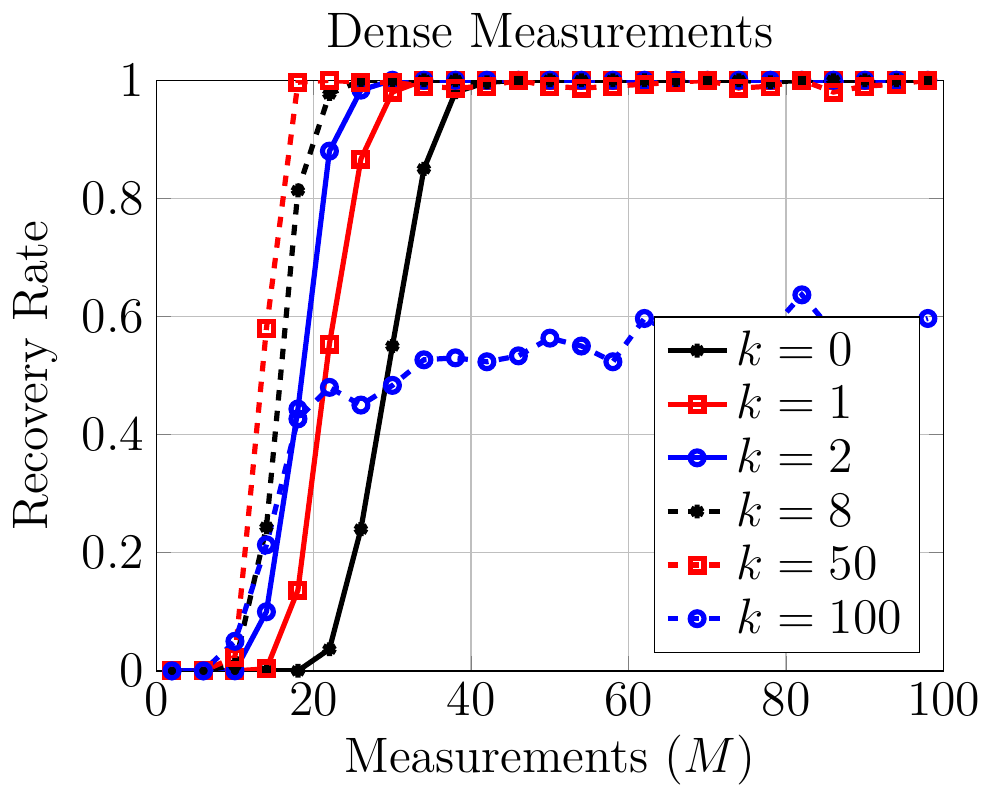}
   \label{fig:NEW_MEAS_TwoDimDiffRecoveryTest2_Tvec_Random_all}
 }
\subfigure[]{
   \includegraphics[width = .45\columnwidth]{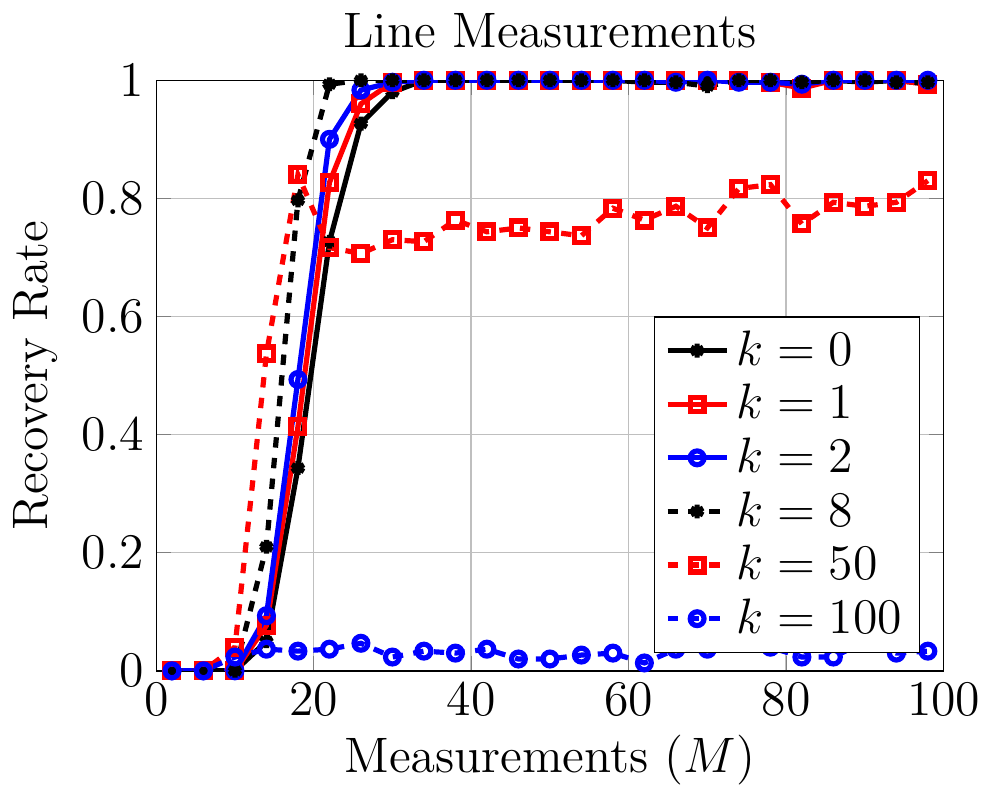}
   \label{fig:NEW_MEAS_TwoDimDiffRecoveryTest2_Tvec_Line_all}
 }
\caption{
Signal recovery from compressive measurements of a diffusion process which has initiated from a sparse initial state of dimension $N = 100$ and sparsity level $S = 9$. The plots show the percent of trials (out of $300$ trials in total) with perfect recovery of the initial state $\vc{x}_0$ versus the number of measurements $M$ taken at observation times $k = \left\{0,1,2,8,50,100\right\}$. (a) Recovery from compressive Dense Measurements. (b) Recovery from compressive Line Measurements.
}
\label{fig:NEW_MEAS_TwoDimDiffRecoveryTest2_Tvec_all}
\end{figure}

In order to see how the recovery performance would change as we take measurements at different times, we repeat the previous example for $k = \left\{0,1,2,8,50,100\right\}$. The results are shown in Fig.~\ref{fig:NEW_MEAS_TwoDimDiffRecoveryTest2_Tvec_Random_all} and Fig.~\ref{fig:NEW_MEAS_TwoDimDiffRecoveryTest2_Tvec_Line_all} for Dense and Line Measurements, respectively. In both cases, the recovery performance starts to improve as we take measurements at later times.
However, in both measuring scenarios, the recovery performance tends to decrease if we wait too long to take measurements. For example, as shown in Fig.~\ref{fig:NEW_MEAS_TwoDimDiffRecoveryTest2_Tvec_Random_all}, the recovery performance is significantly decreased at time $k=100$ when Dense Measurements are employed. A more dramatic decrease in the recovery performance can be observed when Line Measurements are employed in Fig.~\ref{fig:NEW_MEAS_TwoDimDiffRecoveryTest2_Tvec_Line_all}.
Again this behavior is as expected and can be interpreted with the diffusion phenomenon. If we wait too long to take measurements from the field of study (e.g., the lake of water), the effect of the initial contaminant starts to disappear in the field (due to diffusion) and consequently measurements
at later times contain less information. In summary, one could conclude from these observations that taking compressive measurements of a diffusion process at times that are too early or too late might decrease the recovery performance.

\begin{figure}[tb]
\centering
\subfigure[]{
   \includegraphics[width = .45\columnwidth]{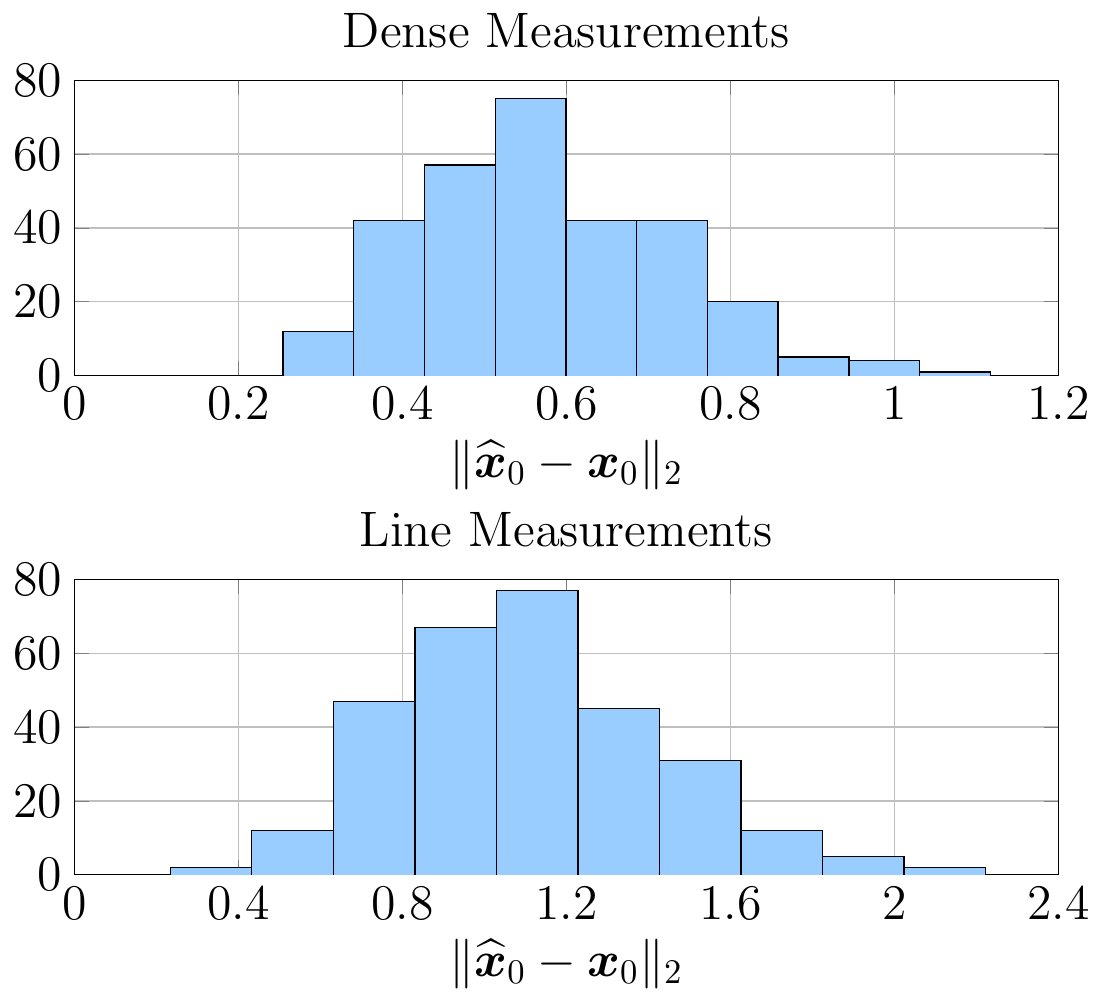}
   \label{fig:NEW_MEAS_TwoDimDiffRecoveryTest2RunNEW_Noise_Tvec_2}
 }
\subfigure[]{
   \includegraphics[width = .45\columnwidth]{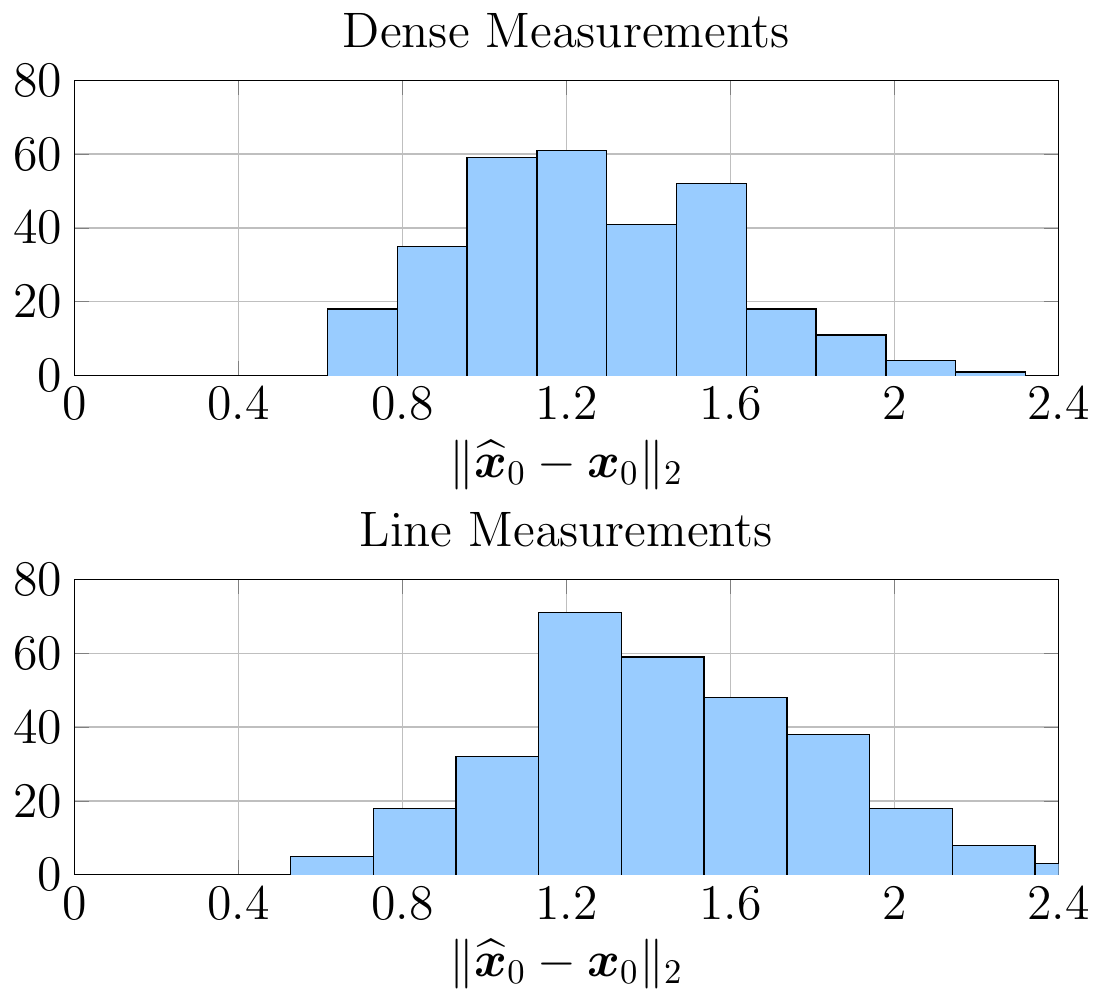}
   \label{fig:NEW_MEAS_TwoDimDiffRecoveryTest2RunNEW_Noise_Tvec_10}
 }
\caption{
Signal recovery from $M=32$ compressive measurements of a diffusion process which has initiated from a sparse initial state of dimension $N = 100$ and sparsity level $S = 9$. The plots show the
recovery error of the initial state $\|\vc{e}\|_2 = \|\widehat{\vc{x}}_0 - \vc{x}_0\|_2$ over $300$ trials.
(a) Recovery from compressive measurements at time $k=2$. (b) Recovery from compressive measurements at time $k=10$.
}
\label{fig:NEW_MEAS_TwoDimDiffRecoveryTest2RunNEW_Noise}
\end{figure}

In another example, we fix $M = 32$, consider the same model for the sparse initial states with $S=9$ as in the previous examples, introduce white noise in the measurements with standard deviation 0.05, use a noise-aware version of the $\ell_1$ recovery algorithm~\cite{CandesRIP}, and plot a histogram of the recovery errors $\| \widehat{\vc{x}}_0 - \vc{x}_0\|_2$.
We perform this experiment at $k=2$ and $k=10$. As can be seen in Fig.~\ref{fig:NEW_MEAS_TwoDimDiffRecoveryTest2RunNEW_Noise_Tvec_2}, at time $k=2$ the Dense Measurements have lower recovery errors (almost half) compared to the Line Measurements. However, if we take measurements at time $k=10$, the recovery error of both measurement processes tends to be similar, as depicted in Fig.~\ref{fig:NEW_MEAS_TwoDimDiffRecoveryTest2RunNEW_Noise_Tvec_10}.

\begin{figure}[tb]
\centering
\subfigure[]{
   \includegraphics[width = 0.45\columnwidth]{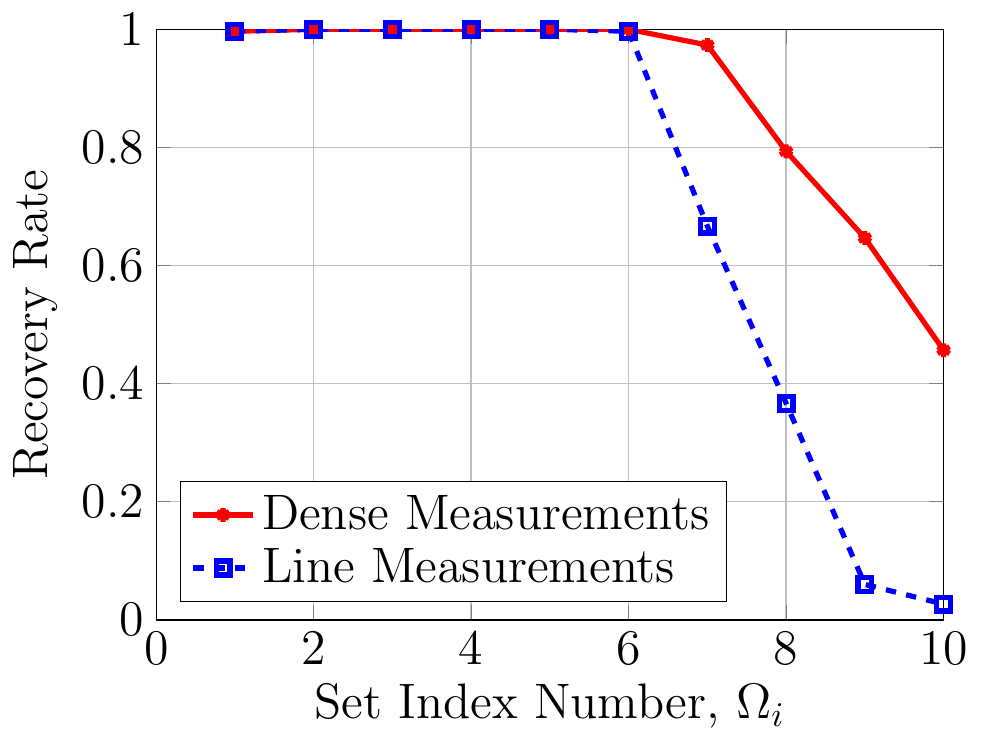}
   \label{fig:NEW_MEAS_TwoDimDiffRecoveryTest3RunNEW_M8}
 }
\subfigure[]{
   \includegraphics[width = 0.39\columnwidth]{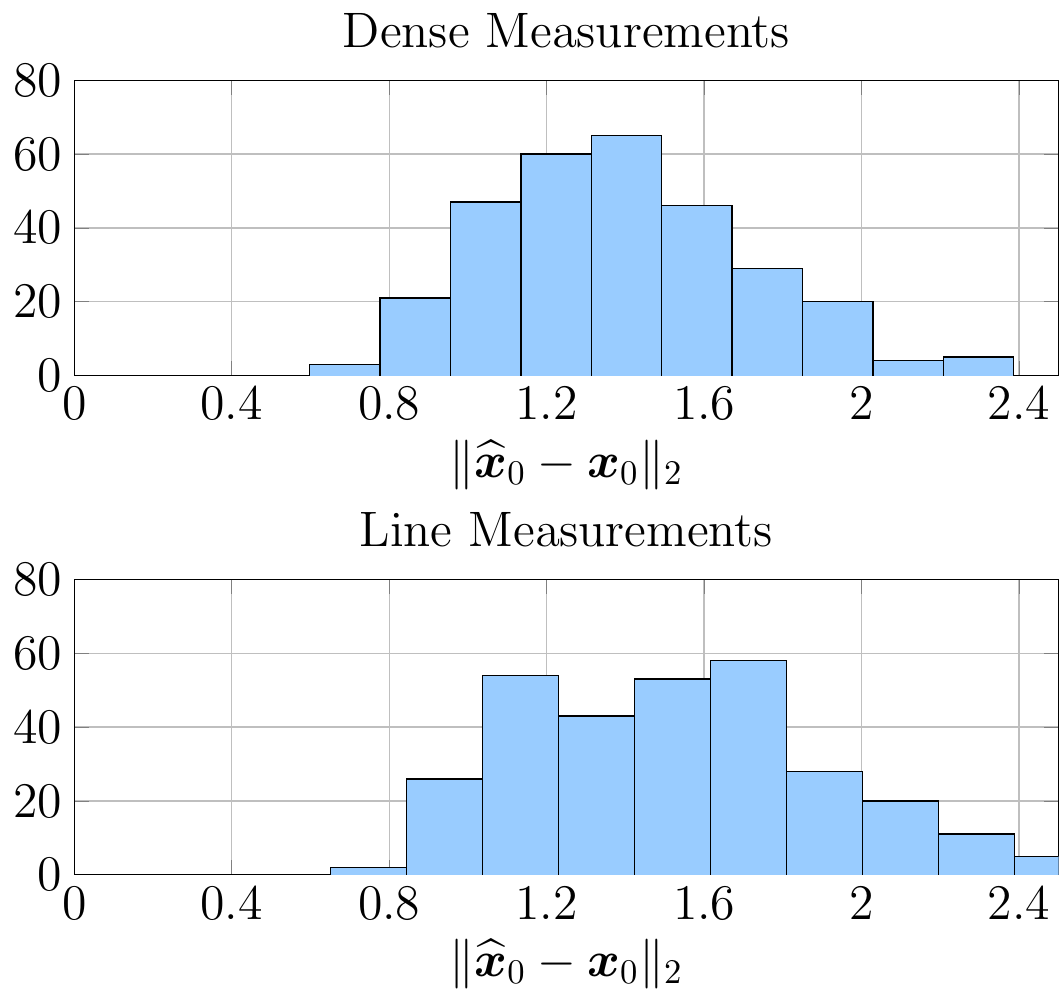}
   \label{fig:NEW_MEAS_TwoDimDiffRecoveryTest3RunNEW_Noise_Tvec_all_M8_eps_20}
 }
\caption{
Signal recovery from compressive measurements of a diffusion process which has initiated from a sparse initial state of dimension $N = 100$ and sparsity level $S = 9$. A total of $KM = 32$ measurements are spread over $K=4$ observation times while at each time, $M=8$ measurements are taken. (a) Percent of trials (out of $300$ trials in total) with perfect recovery of the initial state $\vc{x}_0$ are shown for different sample sets, $\Omega_i$. (b) Recovery error of the initial state $\|\vc{e}\|_2 = \|\widehat{\vc{x}}_0 - \vc{x}_0\|_2$ over $300$ trials for set $\Omega_4$.
}
\end{figure}

Of course, it is not necessary to take all of the measurements only at one observation time. What may not be obvious a priori is how spreading the measurements over time may impact the initial state recovery. To this end, we perform the signal recovery experiments when a total of $MK = 32$ measurements are spread over $K=4$ observation times (at each observation time we take $M=8$ measurements). In order to see how different observation times affect the recovery performance, we repeat the experiment for different sample sets, $\Omega_i$. We consider $10$ sample sets as $\Omega_1 = \left\{0,1,2,3\right\}$, $\Omega_2 = \left\{4,5,6,7\right\}$, $\Omega_3 = \left\{8,9,10,11\right\}$, $\Omega_4 = \left\{10,20,30,40\right\}$, $\Omega_5 = \left\{20,21,22,23\right\}$, $\Omega_6 = \left\{10,30,50,70\right\}$, $\Omega_7 = \left\{51,52,53,54\right\}$, $\Omega_{8} = \left\{60,70,80,90\right\}$, $\Omega_9 = \left\{91,92,93,94\right\}$, and $\Omega_{10} = \left\{97,98,99,100\right\}$.
Figure~\ref{fig:NEW_MEAS_TwoDimDiffRecoveryTest3RunNEW_M8} illustrates the results. For both of the measuring scenarios, the overall recovery performance improves when we take measurements at later times. As mentioned earlier, however, if we wait too long to take measurements the recovery performance drops.
For sample sets $\Omega_2$ through $\Omega_6$, we have perfect recovery of the initial state only from $MK = 32$ total measurements, either using Dense or Line Measurements. The overall recovery performance is not much different compared to, say, taking $M=32$ measurements at a single instant and so there is no significant penalty that one pays by slightly spreading out the measurement collection process in time, as long as a different random measurement matrix is used at each sample time. We repeat the same experiment when the measurements are noisy. We introduce white noise in the measurements with standard deviation 0.05 and use a noise-aware version of the $\ell_1$-minimization problem to recover the true solution.
Figure~\ref{fig:NEW_MEAS_TwoDimDiffRecoveryTest3RunNEW_Noise_Tvec_all_M8_eps_20} depicts a histogram of the recovery errors $\| \widehat{\vc{x}}_0 - \vc{x}_0\|_2$ when $MK = 32$ measurements are spread over $K = 4$ sample times $\Omega_4 = \left\{10,20,30,40\right\}$.

\section*{ACKNOWLEDGMENTS}
The authors gratefully acknowledge Alejandro Weinstein, Armin Eftekhari, Han Lun Yap, Chris Rozell, Kevin Moore, and Kameshwar Poolla for insightful comments and valuable discussions.

\appendix


\subsection{Proof of Theorem~\ref{theo:rip_ok_symmetric_A}}
%
%
We start the analysis by showing that $\|\ak\vc{x}_0\|_2^2$ lies within a small neighborhood around $\|\vc{x}_0\|_2^2$ for any $S$-sparse $\vc{x}_0 \in \real^N$. To this end, we derive the following lemma.

\begin{lemma}
Assume $\Omega = \left\{k_0, k_1, \dots, k_{K-1}\right\}$. Assume $A$ has the eigendecomposition given in~(\ref{eq:decomp1}) and $U_1^T \in \real^{L \times N}$ ($L < N$) satisfies a scaled version of the \ac{RIP} of order $S$ with isometry constant $\delta_S$ as given in~(\ref{eq:rip_U_1^T}). Then, for $\delta_S \in (0,1)$,
\begin{equation}
(1-\delta_S)\frac{L}{N}\sum_{i=0}^{K-1}\lambda_{1,\text{min}}^{2k_i} \leq \frac{\|\ak\vc{x}_0\|_2^2}{\|\vc{x}_0\|_2^2} \leq (1+\delta_S)\frac{L}{N}\sum_{i=0}^{K-1}\lambda_{1,\text{max}}^{2k_i} +  \sum_{i=0}^{K-1}\lambda_{2,\text{max}}^{2k_i}
\label{eq:deterministic_ak}
\end{equation}
holds for all $S$-sparse $\vc{x}_0 \in \real^N$.
\label{lem:deterministic_bnd_ak}
\end{lemma} 

{\textbf{Proof of Lemma~\ref{lem:deterministic_bnd_ak}}}
If $A$ is of the form given in~(\ref{eq:decomp1}), we have
$A\vc{x}_0 = U_1\Lambda_1U_1^T\vc{x}_0 + U_2\Lambda_2U_2^T\vc{x}_0$, and consequently,
\[
\|A\vc{x}_0\|_2^2 = \vc{x}_0^TU_1\Lambda_1^2U_1^T\vc{x}_0+\vc{x}_0^TU_2\Lambda_2^2U_2^T\vc{x}_0 \geq \|\Lambda_1U_1^T\vc{x}_0\|_2^2 \geq \lambda_{1,\text{min}}^2\|U_1^T\vc{x}_0\|_2^2.
\]
On the other hand,
\begin{align*}
\|A\vc{x}_0\|_2^2 = \vc{x}_0^TU_1\Lambda_1^2U_1^T\vc{x}_0+\vc{x}_0^TU_2\Lambda_2^2U_2^T\vc{x}_0 &\leq \lambda_{1,\text{max}}^2\|U_1^T\vc{x}_0\|_2^2 + \lambda_{2,\text{max}}^2\|U_2^T\vc{x}_0\|_2^2 \\
&\leq \lambda_{1,\text{max}}^2\|U_1^T\vc{x}_0\|_2^2 + \lambda_{2,\text{max}}^2\|\vc{x}_0\|_2^2.
\end{align*}
Thus,
\begin{equation}
\lambda_{1,\text{min}}^{2}\|U_1^T\vc{x}_0\|_2^2 \leq \|A\vc{x}_0\|_2^2 \leq \lambda_{1,\text{max}}^{2}\|U_1^T\vc{x}_0\|_2^2 + \lambda_{2,\text{max}}^{2}\|\vc{x}_0\|_2^2.
\label{eq:upper_lower_bnd_A}
\end{equation}
If $U_1^T$ satisfies the scaled \ac{RIP}, then from~(\ref{eq:rip_U_1^T}) and~(\ref{eq:upper_lower_bnd_A})
for $\delta_S \in (0,1)$,
\begin{equation}
(1-\delta_S)\frac{L}{N}\lambda_{1,\text{min}}^{2} \leq \frac{\|A\vc{x}_0\|_2^2}{\|\vc{x}_0\|_2^2} \leq (1+\delta_S)\frac{L}{N}\lambda_{1,\text{max}}^{2} +  \lambda_{2,\text{max}}^{2}
\end{equation}
holds for all $S$-sparse $\vc{x}_0 \in \real^N$. 
Similarly, one can show that for $i \in \left\{0, 1, \dots, K-1\right\}$,
\[
\lambda_{1,\text{min}}^{2k_i}\|U_1^T\vc{x}_0\|_2^2 \leq \|A^{k_i}\vc{x}_0\|_2^2 \leq \lambda_{1,\text{max}}^{2k_i}\|U_1^T\vc{x}_0\|_2^2 + \lambda_{2,\text{max}}^{2k_i}\|\vc{x}_0\|_2^2,
\]
and consequently, for $\delta_S \in (0,1)$,
\begin{equation}
(1-\delta_S)\frac{L}{N}\lambda_{1,\text{min}}^{2k_i} \leq \frac{\|A^{k_i}\vc{x}_0\|_2^2}{\|\vc{x}_0\|_2^2} \leq (1+\delta_S)\frac{L}{N}\lambda_{1,\text{max}}^{2k_i} +  \lambda_{2,\text{max}}^{2k_i}
\label{eq:deterministic_bnd_A_k_i}
\end{equation}
holds for all $S$-sparse $\vc{x}_0 \in \real^N$. Consequently using~(\ref{eq:ak_x_0_sum}), for $\delta_S \in (0,1)$,
\begin{equation*}
(1-\delta_S)\frac{L}{N}\sum_{i=0}^{K-1}\lambda_{1,\text{min}}^{2k_i} \leq \frac{\|\ak\vc{x}_0\|_2^2}{\|\vc{x}_0\|_2^2} \leq (1+\delta_S)\frac{L}{N}\sum_{i=0}^{K-1}\lambda_{1,\text{max}}^{2k_i} +  \sum_{i=0}^{K-1}\lambda_{2,\text{max}}^{2k_i}
\end{equation*}
holds for all $S$-sparse $\vc{x}_0 \in \real^N$. 
\hfill $\blacksquare$

Lemma~\ref{lem:deterministic_bnd_ak} provides deterministic bounds on the ratio $\frac{\|\ak\vc{x}_0\|_2^2}{\|\vc{x}_0\|_2^2}$ for all $S$-sparse $\vc{x}_0$ when $U_1^T$ satisfies the scaled \ac{RIP}. Using this deterministic result, we can now state the proof of Theorem~\ref{theo:rip_ok_symmetric_A} where we show that a scaled version of $\ok$ satisfies the \ac{RIP} with high probability. 

%
First observe that when all matrices $C_k$ are independent and populated with \ac{i.i.d.} Gaussian random entries, from Corollary~\ref{cor:obs_indep} we have the following \ac{CoM} inequality for 
$\ck$. For any fixed $S$-sparse $\vc{x}_0 \in \real^N$, let $\vc{v} = \ak\vc{x}_0 \in \real^{NK}$. Then for any $\epsilon \in (0,\frac{16}{\sqrt{K}})$,
\begin{equation}
\Prob{\bigg{|}\|\ck \vc{v} \|_{2}^{2}- \|\vc{v}\|_2^2\bigg{|} > \epsilon  \|\vc{v}\|_{2}^{2}}
\leq 2\exp\left\{-\frac{M\Gamma(\vc{v})\epsilon^{2}}{256}\right\}.
\label{eq:CoM_signal_dependent}
\end{equation}
As can be seen, the right-hand side of (\ref{eq:CoM_signal_dependent}) is signal dependent. However, we need a universal failure probability bound (that is independent of $\vc{x}_0$) in order to prove the \ac{RIP} based a \ac{CoM} inequality. Define
\begin{equation}
\rho := \inf_{S-\text{sparse}~\vc{x}_0 \in \real^N} \Gamma(\ak\vc{x}_0).
\label{eq:def_min_Gamma}
\end{equation}
Therefore from (\ref{eq:CoM_signal_dependent}) and (\ref{eq:def_min_Gamma}), for any fixed $S$-sparse $\vc{x}_0 \in \real^N$ and for any $\epsilon \in (0,\frac{16}{\sqrt{K}})$,
\begin{equation}
\Prob{\bigg{|}\|\ck \ak\vc{x}_0 \|_{2}^{2}- \|\ak\vc{x}_0\|_2^2\bigg{|} > \epsilon  \|\ak\vc{x}_0\|_{2}^{2}}
\leq 2\exp\left\{-\frac{M\rho\epsilon^{2}}{256}\right\}
= 2\exp\left\{-\km f(\epsilon)\right\},
\label{eq:CoM_ck_sparse}
\end{equation}
where $f(\epsilon) := \frac{\rho\epsilon^{2}}{256K}$, $\km := MK$, and $\kn := NK$. Let $\nu \in (0,1)$ denote a failure probability and $\delta \in (0,\frac{16}{\sqrt{K}})$ denote a distortion factor.
Through a union bound argument and by applying Lemma~\ref{lem:RIP_based_CoM} for all $N \choose S$ $S$-dimensional subspaces in $\real^N$, whenever $\ck \in \real^{\km \times \kn}$ satisfies the \ac{CoM} inequality~(\ref{eq:CoM_ck_sparse}) with
\begin{equation} 
MK \geq \frac{S\log(\frac{42}{\delta})+\log(\frac{2}{\nu})+\log({N \choose S})}{f(\frac{\delta}{\sqrt{2}})},
\end{equation}
then with probability exceeding $1-\nu$,
\[
(1-\delta)\|\ak\vc{x}_0\|_2^2 \leq \|\ck\ak\vc{x}_0\|_2^2 \leq (1+\delta)\|\ak\vc{x}_0\|_2^2,
\]
for all $S$-sparse $\vc{x}_0 \in \real^N$. Consequently using the deterministic bound on $\|\ak\vc{x}_0\|_2^2$ derived in~(\ref{eq:deterministic_ak}), with probability exceeding $1-\nu$,
\[
(1-\delta)\left((1-\delta_S)\frac{L}{N}\sum_{i=0}^{K-1}\lambda_{1,\text{min}}^{2k_i}\right) \leq \frac{\|\ok\vc{x}_0\|_2^2}{\|\vc{x}_0\|_2^2} \leq (1+\delta)\left((1+\delta_S)\frac{L}{N}\sum_{i=0}^{K-1}\lambda_{1,\text{max}}^{2k_i} +  \sum_{i=0}^{K-1}\lambda_{2,\text{max}}^{2k_i}\right)
\]
for all $S$-sparse $\vc{x}_0 \in \real^N$. 
\hfill $\blacksquare$


\subsection{Proof of Corollary~\ref{cor:rip_ok_symmetric_A_extreme} and Corollary~\ref{cor:rip_ok_symmetric_A_extreme_1}}

We simply need to derive a lower bound on $\Gamma(\ak\vc{x}_0)$ as an evaluation of $\rho$. Recall~(\ref{eq:gamma1}) and define
\[
\vc{z}_0 := \left[\|A^{k_0}\vc{x}_0\|_2^2 \ \|A^{k_1}\vc{x}_{0}\|_2^2 \ \cdots \ \|A^{k_{K-1}} \vc{x}_0\|_2^2\right]^T \in \real^K.
\]
If all the entries of $\vc{z}_0$ lie within some bounds as $\ell_{\ell} \leq z_0\left(i\right) \leq \ell_h$ for all $i$, then one can show that
\begin{equation}
\Gamma(\ak \vc{x}_0) \geq K\left(\frac{\ell_{\ell}}{\ell_h}\right)^2.
\label{eq:upper_bnd_Gamma_1}
\end{equation}
Using the deterministic bound derived in~(\ref{eq:deterministic_bnd_A_k_i}) on $\|A^{k_i}\vc{x}_0\|_2^2$ for all $i \in \left\{0,1,\dots,K-1\right\}$, one can show that when $\lambda = 1$ ($\lambda_{1,\text{max}} = \lambda_{1,\text{min}} = \lambda$ and $\lambda_{2,\text{max}} = 0$), $\ell_{\ell} = (1-\delta_S)\frac{L}{N}\|\vc{x}_0\|_2^2$ and $\ell_{h} = (1+\delta_S)\frac{L}{N}\|\vc{x}_0\|_2^2$, and thus,
\[
\rho \geq K\frac{(1-\delta_S)^2}{(1+\delta_S)^2}.
\]
Similarly one can show that when $\lambda <1$, 
\begin{equation}
\rho \geq K\frac{(1-\delta_S)^2}{(1+\delta_S)^2}\lambda^{4(k_{K-1}-k_0)},\label{eq:Gamma_bnd_lambda_L1}
\end{equation}
and when $\lambda >1$,
\begin{equation}
\rho \geq K\frac{(1-\delta_S)^2}{(1+\delta_S)^2}\lambda^{-4(k_{K-1}-k_0)}.\label{eq:Gamma_bnd_lambda_G1}
\end{equation}
Using these lower bounds on $\rho$ (recall that $\rho$ is defined in (\ref{eq:def_min_Gamma}) as the infimum of $\Gamma(\ak \vc{x}_0)$ over all $S$-sparse $\vc{x}_0 \in \real^N$) in the result of Theorem~\ref{theo:rip_ok_symmetric_A} completes the proof.
We also note that when $\lambda_{1,\text{max}} = \lambda_{1,\text{min}} = \lambda$ and $\lambda_{2,\text{max}} = 0$, the upper bound given in~(\ref{eq:CoM_ck_sparse}) can be used to bound the left-hand side failure probability even when $\epsilon \geq \frac{16}{\sqrt{K}}$. In fact, we can show that~(\ref{eq:CoM_ck_sparse}) holds for 
any $\epsilon \in (0,1)$. The \ac{RIP} results of Corollaries~\ref{cor:rip_ok_symmetric_A_extreme} and
\ref{cor:rip_ok_symmetric_A_extreme_1} follow based on this \ac{CoM} inequality. We have omitted these details for the sake of space.


\footnotesize
\bibliographystyle{IEEEbib}
\bibliography{referfile}

\end{document}